\documentclass[12pt]{article}
\pdfoutput =1
\usepackage{graphics}
\usepackage{graphicx}
\DeclareGraphicsExtensions{.pdf}
\usepackage{float} %Include figure filesusepackage{graphicx} %Include figure files
\usepackage{subfloat}
\usepackage{amsmath}
\usepackage{slashed}
\usepackage{amsfonts}  
\usepackage{amssymb}

\def\b{\beta}

\def\s{\sigma}

\def\th{\theta}

\def\Om{\Omega}

\usepackage{comment}
\usepackage{color}

\usepackage{subcaption}
\begin{document}
\date{}
%%%%%%%%%%%%%%%%%%%%
\title{{\bf{\Large Metallic transports from Taub-NUT AdS black holes}}}
%%%%%%%%%%%%%%%%%%%%
\author{
{\bf { ~Mohd Aariyan Khan}$
$\thanks{E-mail:  aariyan\_km@ph.iitr.ac.in}}\\
 {\normalsize  Department of Physics, Indian Institute of Technology Roorkee,}\\
  {\normalsize Roorkee 247667, Uttarakhand, India}
\\[0.3cm]
{\bf { Hemant Rathi}$
$\thanks{E-mail:  hemant.rathi@saha.ac.in}}\\
 {\normalsize  Saha Institute of Nuclear Physics, 1/AF Bidhannagar, }\\
  {\normalsize Kolkata 700064, India }
  \\[0.3cm]
 {\bf { ~Dibakar Roychowdhury}$
$\thanks{E-mail:  dibakar.roychowdhury@ph.iitr.ac.in}}\\
 {\normalsize  Department of Physics, Indian Institute of Technology Roorkee,}\\
  {\normalsize Roorkee 247667, Uttarakhand, India}
\\[0.3cm]
}
\maketitle
%%%%%%%%%%%%%%%%%%%%%%%%%%%%%%%%%%%%%%%%%%%%%%%%%%%%%%%%%%%%
\abstract{We compute holographic DC conductivity associated with the Taub-NUT-\(AdS_4\) black holes following the probe D-brane approach. In particular, we examine the effects of frame dragging on charge transport in both low and high temperature regimes. Our analysis reveals that in the low temperature regime, the conductivity is sensitive to the presence of the Misner string that causes frame dragging. Notably, the increase in the conductivity near the Misner string is sharper as compared to points farther away from it. On the other hand, in the high temperature regime, the effects due to frame dragging are significantly suppressed, and the thermal contribution to the charge transport takes over that due to the U(1) charge carriers.}
%%%%%%%%%%%%%%%%%%%%%%%%%%%%%%%%%%%%
\section{Overview and motivation }

The Taub-NUT (TN) space-time is a solution of Einstein's equations in four dimensions \cite{Taub:1950ez}. It was initially discovered by Taub in 1951 and addresses only the time dependent part of the full space-time. Later, in 1963, Newman, Tamburino, and Unti rediscovered it as a straightforward generalization of the Schwarzschild space-time in the presence of the NUT parameter $n$ \cite{Newman:1963yy}. However, the TN space-time does not reduce to standard Minkowski space-time in the asymptotic limit. A notable feature of the Taub-NUT metric is the presence of the Misner string, which leads to the non-trivial frame dragging effects\footnote{We will explore the TN space-time in more detail in Section \ref{ri}.} \cite{Durka:2019ajz}-\cite{Ong:2016cbo}.

%\textcolor{red}{The authors in\cite{Bordo:2019tyh} -\cite{Rodriguez:2021hks} discuses the thermodynamics of TN(AdS) black hole in different theories. In particular }

The authors in \cite{Bordo:2019tyh}-\cite{Rodriguez:2021hks} further generalized Taub-NUT solutions for the Anti-de Sitter (AdS) space-time in four dimensions and investigated the thermodynamics of the Taub-NUT AdS (TN-AdS) black hole. In order to satisfy the first law of black hole thermodynamics, the authors in \cite{Bordo:2019tyh} introduced the notion of the Misner potential $(\psi)$ and the Misner charge ($N$), and treated the parameter $m$ as the total mass of the TN-AdS black hole. Their analysis suggested that the entropy of the TN-AdS black hole is significantly affected due to the presence of the NUT parameter $(n)$.

Later on, the authors in \cite{Liu:2023uqf} re-investigated the thermal properties of the TN-AdS black holes. They demonstrated that the NUT parameter $(n)$ acts like a thermodynamic potential ($\Phi_n$), which is conjugate to the NUT charge \(Q_n\)  and the \(Q_n\) is distributed over the Misner string. Furthermore, they showed that the NUT parameter $(n)$ significantly affects both the entropy and total mass, $M=m+2\Phi_nQ_n$, of the TN-AdS black hole.

% Furthermore, the authors in \cite{Liu:2023uqf} demonstrated that the NUT parameter $(n)$ acts as a thermodynamic potential ($\Phi_n$), which is conjugate to the NUT charge \(Q_n\)  and the \(Q_n\) is distributed over the Misner string.

%\cite{Liu:2023uqf} 

Another interesting property of these solutions has been investigated by the authors in \cite{Jiang:2019yzs}-\cite{Chen:2023eio}. In particular, they study the holographic complexity of a charged Taub-NUT AdS black hole. Their analysis reveals that the complexity depends on both the Misner potential $(\psi)$ and the Misner charge ($N$). As a result, the complexity growth rate at late times is affected due to the presence of the Misner string.

%\textcolor{red}{Earlier, an Author in \cite{Bordo:2019tyh} satisfied the thermodynamics 1st Law by treating m parameter as the total mass of the TN-Ads black hole, in which he introduced Misner's potential $\psi$ and Misner charge $N$ associated with Misner's string additionally it can be observed easily by Area/4 law that entropy is modified due to nut parameter.and An Author in \cite{Jiang:2019yzs} computed the holographic complexity of a charged-Taub-NUT-Ads black hole and showed that Misner's string affects the late-time complexity growth rate as its depends on Misner's potential $\psi$ and Misner's charge $N$.
%Recently, the authors in \cite{Liu:2022wku} re-investigated the thermodynamics of the Taub-NUT  black hole and generalised for Taub-NUT-Ads black hole in \cite{Liu:2023uqf},in which they found that the total mass of the black hole is not the m parameter but  $M=m+2\Phi_n Q_n$ that solve a very serious problem of the Taub-NUT black hole that it also exists for -ve m, but with the identification of total mass of the black hole by M , total mass of the black hole is always +ve since M is always +ve even for -ve m ,they also setify the 1st law of thermodynamics by treating M as the mass of the TN(Ads) black hole. Furthermore, the authors in \cite{Liu:2023uqf} demonstrated that the NUT parameter $n$ acts as a thermodynamic potential$\Phi_n$, which is conjugate to the NUT charge \(Q_N\)  and the \(Q_N\) is distributed over the Misner string.} 

Recently, the authors in \cite{Kalamakis:2020aaj}-\cite{Kalamakis:2025daq} have explored the TN-AdS space-time in the context of AdS/CMT. In particular, the authors in \cite{Kalamakis:2020aaj} examined the quasi-normal modes of a scalar field propagating over the TN-AdS background. Their findings indicated that if the Misner string is considered invisible, then TN-AdS space-time is dual to a holographic non-dissipative superfluid with quantized-vortex excitations. On the other hand, if the Misner string is treated as a physical entity, then TN-AdS corresponds to a holographic dissipative fluid, and its properties are typically fixed through the complex quasi-normal modes of bulk fluctuations.

%Despite the interesting feature of TN AdS space-time in the Misner string, this space-time has been explored less in the context of holography. \cite{kalamakis2021aspects} The text examines the quasi-normal modes of a scalar field within a TNAds background. This demonstrates that if Misner's string is considered invisible, TNAds spacetime represents a holographic non-dissipative superfluid with quantized-vortex excitations. Conversely, if the Misner string is regarded as a physical entity, then TNAdS4 is akin to a holographic fluid, and its dissipative characteristics are typically explored through complex quasinormal modes of bulk fluctuations.

Despite several interesting studies on TN-AdS space-time as mentioned above, the metallic transport properties of the dual QFT have not been explored yet. The purpose of this paper is to fill this gap. In particular, we carry out a detailed analysis and build up the notion of metallic holography for the Taub-NUT black hole along the lines of \cite{Karch:2007pd}-\cite{Hartnoll:2009ns}.

In the present paper, we compute the holographic DC conductivity associated with the TN-AdS black hole and, in particular, examine the effects of the NUT parameter ($n$) on it. We carry out analysis both in the low and the high temperature regimes. Finally, we discuss the interplay between the conductivity at low temperatures and that of a Fermi liquid \cite{Ge:2015fmu}-\cite{Lee:2010uy}.

Following the approach as outlined by the authors in \cite{Karch:2007pd}, we analyze the TN-AdS black hole using a probe D-brane, which sources massless U(1) charge carriers for the dual QFT living at the boundary. We apply an external electric field ($E$) that causes these charge carriers to drift, resulting in a current flowing in the direction of the electric field. Finally, we compute the conductivity associated with these charge carriers, treating the external electric field $E$ as a small parameter, i.e., $E<<1$. 

Our analysis reveals that the total conductivity is composed of two parts. The first component ($\s_{U(1)}$) arises due to the explicitly added U(1) charge carriers, while the other component is sourced due to the thermally produced charge pairs, which we denote as $\s_{thermal}$.

At low temperatures, the number of electric charge carriers dominates over the thermally produced carriers, leading to a higher U(1) conductivity. On the other hand, at high temperatures, the thermally produced charge pairs exceed the population of U(1) charge carriers. As a result, the thermal conductivity dominates over the U(1) conductivity, $\s_{thermal}>>\s_{U(1)}$.

%At low temperatures, we observed that the conductivity due to the U(1) charge carriers increases as we approach the threshold temperature, regardless of the position of the Misner string. However, the thermal conductivity rises as we move close to the minimum temperature only in the vicinity of the Misner string, while it remains constant with temperature at points farther away.

%In addition, we found that the increase in conductivities ($\s_{U(1)}$ and $\s_{thermal}$) occurs more sharply near the Misner string as compared to the more distant locations. This phenomenon arises due to the presence of the NUT parameter, which causes novel frame dragging near the Misner string. As a result, the U(1) charge carriers and the thermally produced charge pairs experience greater drift due to this frame dragging, leading to higher conductivities near the Misner string.

%On the other hand, the effects of frame dragging are diminished at higher temperatures, resulting in similar U(1) conductivity behavior both near and far from the Misner string. This observation also applies to the thermal conductivity.

As a notable feature of our analysis, we notice that at low temperatures, both $\s_{U(1)}$ and $\s_{thermal}$ are largely affected due to the frame dragging effects near the Misner string. We identify this as a physical effect that produces an extra drift to both types of charge carriers near the Misner string, causing an increase in the conductivity. However, as we move towards higher temperatures, the effects due to frame dragging are largely suppressed. In other words, both the conductivities $\s_{U(1)}$ and $\s_{thermal}$ remain unaffected as we approach the Misner string. 

The rest of the paper is organized as follows

$\bullet$ In section \ref{ri}, we review the Taub-NUT space-time in four dimensions and discuss novel frame dragging effects associated with it.
 
$\bullet$ In Section \ref{tnbody}, we review Taub-NUT black hole solutions for AdS space-time and compute the Hawking temperature associated with it. Furthermore, we introduce probe D-branes and construct the associated Dirac-Born-Infeld (DBI) action.
  
$\bullet$  In Section \ref{DC conductivity}, we compute the holographic DC conductivity associated with the TN-AdS black holes, treating the external electric field ($E$) as a perturbation (\(E<<1\)).
 
$\bullet$  In Section \ref{nuteffectsec}, we examine the effect of the NUT parameter ($n$) and associated frame dragging on the holographic DC conductivity in both low and high temperature regimes. 

$\bullet$ In Section \ref{cncbody}, we conclude our discussion and highlight several interesting research directions that are worth pursuing in future.

\section{Taub-NUT space-time in nutshell}\label{ri}
In this Section, we briefly review the Taub-NUT space-time \cite{Newman:1963yy} and explore the novel frame dragging \cite{Durka:2019ajz}-\cite{Ong:2016cbo} associated with it. In the next Section, we generalize these solutions for AdS space-time, which will set the stage for our computation.
\subsection{General introduction} \label{sub1}
The Taub-NUT space-time is a stationary, axisymmetric solution of the four-dimensional Einstein's equation \cite{Taub:1950ez}. The TN metric in four dimensions can be expressed as \cite{Newman:1963yy}
\begin{align}
    ds_{TN}^{2}=&\hspace{1mm}- f(r)(dt-2n\cos\theta d\phi)^{2}+\frac{1}{f(r)}dr^{2}+(r^{2}+n^2) (d\theta^{2}+\sin^2 \theta d\phi^{2}),\nonumber\\
   f(r)=&\hspace{1mm}1-\frac{2(mr+n^2)}{r^{2}+n^{2}}, \label{TN metric in r}
\end{align}
where $m$ is the mass parameter and $n$ corresponds to the NUT  parameter\footnote{In literature, the analogy between the NUT solution and Dirac’s theory of magnetic monopoles has been discussed \cite{Dowker:1974znr}. The NUT parameter $n$ is considered as the gravitational analog of the magnetic monopole's ``magnetic mass".}.

It is worth noting that the above equation (\ref{TN metric in r}) reduces to the well-known four dimensional Schwarzschild metric in the limit $n\rightarrow0$. However, it does not reduce to the flat space-time metric in the asymptotic limit ($r\rightarrow \infty$), as given below
\begin{align}
     ds^2_{TN}\Big|_{r\rightarrow\infty}\approx- (dt-2n\cos\theta d\phi)^{2}+dr^{2}+r^{2}(d\theta^{2}+\sin^2\theta d\phi^{2}).  \label{ATN}
\end{align}
Instead, the TN metric is locally flat in the asymptotic limit, which means that all the components of the Riemann curvature tensor (i.e. $R_{\mu \nu \rho \sigma}\sim r^{-3})$ becomes negligibly small as we approach \(r\rightarrow \infty\) \cite{Misner:1963fr}.

In literature \cite{Misner:1963fr}, the TN metric is commonly expressed in the ($t,x,y,z$) coordinates as follows
\begin{align}
  ds_{TN}^{2}=- f(r)\left[dt-2n\frac{z}{r}\frac{(xdy-ydx)}{(r^2-z^2)}\right]^{2}+\frac{1}{f(r)}dr^{2}+\frac{(r^{2}+n^{2})}{r^2}(dx^2+dy^2+dz^2-dr^2) ,\label{p2} 
\end{align}
where $r=\sqrt{x^2+y^2+z^2}$. It is important to note that, in this representation (\ref{p2}), the TN metric features a line singularity at $r=\pm z$ (or at $\th=0,\pi$), which is referred to as the Misner string \cite{Misner:1963fr}. 

In order to understand the Misner string, the authors in   \cite{Misner:1963fr} consider these line singularities as coordinate singularities and propose a coordinate transformation to remove them. In particular, the authors divide space-time into two patches. In the first patch, where \(0\leq \theta <\pi\), \(t\) transformed as \(t + 2n\phi\), and in the second region patch, where \(0< \theta \leq \pi\), \(t\) transformed as \(t - 2n\phi\). This approach removes the line singularity, but time becomes periodic everywhere with a periodicity of \(8 \pi n\).

An alternative approach was proposed by the authors in \cite{Bonnor:1969ala}. In this method, the authors eliminated the singularity at \(\theta=0\) using a coordinate transformation, \(t\rightarrow t + 2n\phi\) for complete space-time, and considered the line singularity at \(\theta=\pi\) as a physical singularity and a massless source of angular momentum.

It is worth noting that, apparently, it looks like the TN space-time contains a closed timelike curve. To demonstrate this fact, we treat the variables $t,r$, and $\th$ in the equation (\ref{TN metric in r}) as constants. This yields the following result
\begin{align}
    ds^{2}=-\sin^2(\theta_0)\left[4n^2f(r_0)\tan^2\left(\frac{\theta_0}{2}\right)-(r_0^2+n^2)\right]d\phi^2,
\end{align}
where $\th_0, r_0$ and $t_0$ are constants. 

Notice that the variable \(\phi\) has a periodic nature. In the region where \(\theta_0 > 2\tan^{-1}\left(\sqrt{\frac{r_0^2+n^2}{4n^2f(r_0)}}\right)\), \(\phi\) becomes timelike, which can lead to the formation of closed timelike curves. However, the authors in \cite{Clement:2015cxa} conduct a thorough analysis of geodesics and show that the TN spacetime does not contain closed timelike curves. Therefore, there will be no constraints on the variable \(\theta\). %. These curves are problematic because they violate causality. However, one can prevent them by restricting the range of the variable \(\theta\) to \(0 \leq \theta < 2\tan^{-1}\left(\sqrt{\frac{r_0^2+n^2}{4n^2f(r_0)}}\right)\) *\footnotemark[2].

\subsection{Frame dragging}
Frame dragging is a fundamental concept in the context of general relativity. It occurs when a massive object, such as a black hole, rotates. This rotation causes the surrounding space-time to be twisted and dragged along with it. As a result, any test particle located near the black hole is influenced by this rotation, and the associated angular velocity can be expressed as \cite{Zhang:2016gzk}
\begin{align}\label{omegadef}
    \Omega=-\frac{g_{t\phi}}{g_{\phi \phi}}.
\end{align}

A well-known example of a black hole that demonstrates the frame dragging in space-time is the Kerr black hole \cite{Ong:2016cbo}. The Kerr metric describes the space-time outside a rotating black hole. The angular velocity of rotation of space-time is given by \cite{Ong:2016cbo}
\begin{align}
    \Omega=-\frac{g_{t\phi}}{g_{\phi \phi}}=\frac{2M r \alpha}{\rho^2(r^2+\alpha^2)+2Mr\alpha^2\sin^2\theta},
\end{align}
where $r$ is the radial coordinate, $M$ is the mass, $J=\alpha M$ is the angular momentum and $\rho^2=r^2+\alpha^2\cos^2\theta$.

Notice that the TN metric (\ref{TN metric in r})  represents the space-time around a ``non-rotating" black hole. However, the NUT parameter ($n$) introduces a frame dragging , and the rotation comes from Misner's string \cite{Durka:2019ajz}. This aligns with  Bonnor's interpretation of the TN metric \cite{Bonnor:1969ala}, which represents the field of spherically symmetric mass with a semi-infinite massless source of angular momentum along the axis of symmetry. Consequently, the angular velocity associated with this rotation depends both on the NUT parameter as well as the position of the Misner string. 

\begin{figure}[H]
    \centering
    \includegraphics[width=0.8\linewidth]{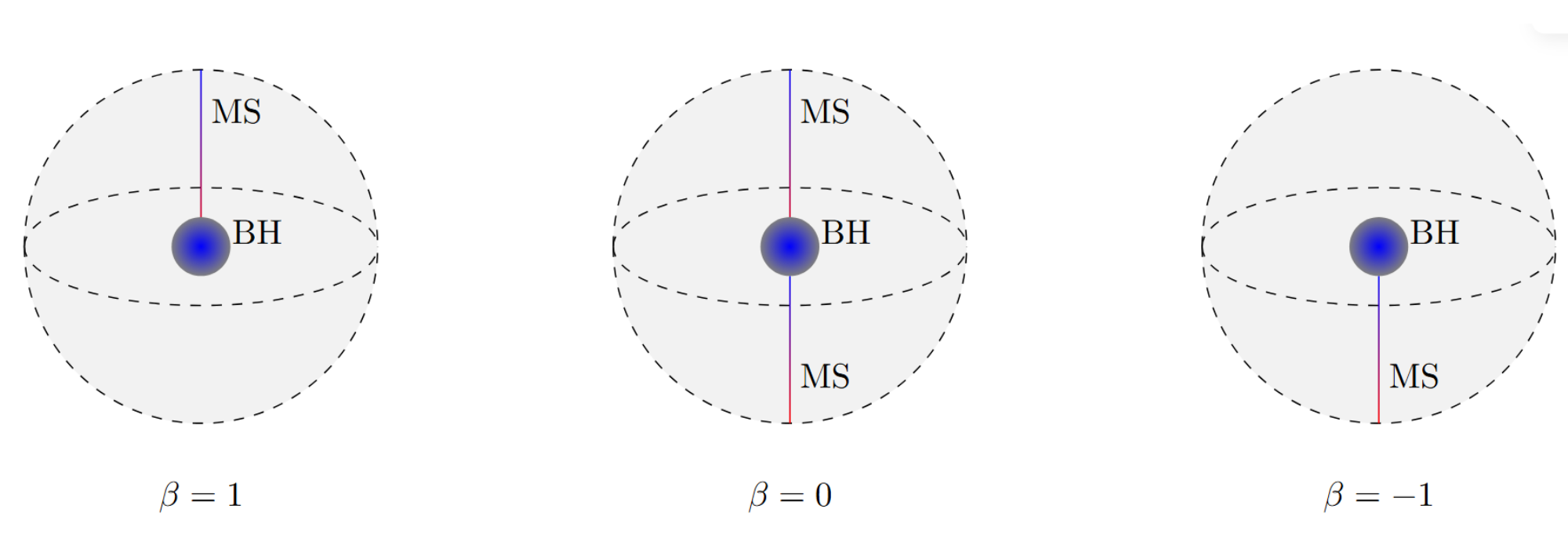}
    \caption{Position of the Misner string for different values of $\b$. Here, BH denotes the black hole and MS denotes Misner's string.}
    \label{fig:placeholder}
\end{figure}

In order to explicitly define the position of the Misner string in the TN metric (\ref{TN metric in r}), we introduce a parameter \(\beta\). This leads to the following expression \cite{Durka:2019ajz}
\begin{align}
     ds_{TN}^{2}=- f(r)\left(dt-2n \left(\cos\theta+\beta\right) d\phi\right)^{2}+\frac{1}{f(r)}dr^{2}+(r^{2}+n^2) (d\theta^{2}+\sin^2 \theta d\phi^{2}), \label{TN metric in r,gen MS}
\end{align}
where $\beta$ determines the position of the Misner string and can take the values $0,\pm1$. For $\beta=0$, the Misner string is located at either \(\theta=0\) or \(\theta=\pi\). However, for $\beta=1$ and $\beta=-1$, the Misner string is located only at $\theta=0$ and $\theta=\pi$, respectively\footnote{It is worth noting that the authors in \cite{Clement:2015cxa} demonstrated that the TN space-time does not have any closed timelike curves for $|\b|\leq1$. Therefore, there are no restrictions on the variable \( \th \).%\textcolor{red}{It can apparently look like there are closed timelike curves for} $\left(\frac{\cos\theta+\beta}{\sin\theta}\right)^2>\frac{(r_0^2+n^2)}{4n^2f(r_0)}$,To avoid closed timelike curves, \(\theta\) must lie in the range \(\cot^2\left(\frac{\theta}{2}\right)<\frac{(r^2+n^2)}{4n^2f(r)}\) for \(\beta=1\). For \(\beta=0\), the range is \(\cot^2\theta<\frac{(r^2+n^2)}{4n^2 f(r)}\) and for \(\beta=-1\), the condition is \(\tan^2\left(\frac{\theta}{2}\right)<\frac{(r^2+n^2)}{4n^2f(r)}\). but Gérard Clément did complete analysis of geodesics and show that there are no closed timelike curves for $|\beta|\leq1$ hance there will be no constraint on $\theta$\cite{Clement:2015cxa}\\ 
}, as shown in the Figure \ref{fig:placeholder}.

Using (\ref{omegadef}) and (\ref{TN metric in r,gen MS}), one can express the angular velocity of the frame dragging as 
\begin{equation}
    \Omega=-\frac{g_{t \phi}}{g_{\phi \phi}}=\frac{-2n f(r) \left(\cos\theta +\beta\right)}{   \left(r^2+n^2\right)\sin^2\theta-4n^2f(r) \left(\cos\theta +\beta\right)^2 } \label{frame dragging Omega}.
\end{equation}

It is important to note that the angular velocity (\ref{frame dragging Omega}) is directly proportional to \( f(r) \). As a result, the effects of frame dragging completely disappear at the horizon (\( r_h \)), where \( f(r_h) = 0 \), regardless of the configuration of the Misner string.

\section{Taub-NUT AdS black hole and DBI action}\label{tnbody}
In this section, we discuss the Taub-NUT generalization of standard Anti-de Sitter black holes and the associated Hawking temperature. We use these solutions, introduce the Dirac-Born-Infeld (DBI) action \cite{Karch:2007pd}, and outline the computation for the holographic DC conductivity.

The space-time metric of the TN-AdS black hole in four dimensions \cite{Durka:2019ajz} can be expressed as
   \begin{align}
         ds^{2}_{TNAdS}=-f(r)\left[dt-2n (\cos\theta+\beta)d\phi\right]^2+f(r)^{-1}dr^{2}+(r^{2}+n^{2})d\Omega_{2}^{2}(\theta,\phi), \label{TN ads}
    \end{align}
    where we denote
     \begin{align}
    f(r)=\frac{r^{2}-n^{2}-2mr+L^{-2}(r^{4}+6n^{2}r^{2}-3n^{4})}{r^{2}+n^{2}}, 
\end{align}
and $d\Omega_{2}^{2}(\theta,\phi)=d\theta^2+\sin^2\theta d\phi^2$. 

The Hawking temperature associated with the TN-AdS black hole can be expressed as 
\begin{align}\label{htem}
T=\frac{f'(r_h)}{4 \pi}=\frac{1}{4 \pi r_h}\left[1+\frac{3(n^2+r_h^2)}{L^2}\right],
\end{align}
where $r_h$ is the event horizon that satisfies $f(r)\Big|_{r=r_h}=0$.

Notice that the boundary of TN-AdS space-time is located at $r\rightarrow \infty$ and in the large $r$ limit, the line element (\ref{TN ads}) takes the following form
\begin{align}
    ds^2_{TNAdS}\Big|_{r\rightarrow \infty}\approx-\frac{r^2}{L^2}(dt-2n (\cos\theta+\beta) d\phi)^2+\frac{L^2}{r^2}dr^2+r^2d\Omega_2^2(\theta,\phi), \label{asymptotic TAds}
\end{align}
which is clearly not an asymptotically AdS. However, on closer inspection, one finds that all the components of the Riemann tensor for the TN-AdS black hole are identical to those of the AdS black hole in the limit $r\rightarrow \infty$, which means that TN-AdS is locally AdS in the asymptotic regime \cite{TN and M}.

For the present analysis, we rewrite the TN-AdS space-time (\ref{TN ads}) in the $z$ coordinate using the coordinate transformation $z=\frac{L^2}{r},$ which yields
 \begin{align}
         ds^{2}=\hspace{1mm}&-f(z)\left[dt^2+4n^2(\cos\theta+\beta)^2d\phi^2 -4n(\cos\theta+\beta)d\phi dt\right]+\frac{L^4}{z^4f(z)}dz^{2}+\nonumber\\
         &\frac{(L^4+n^{2}z^2)}{z^2}d\Omega_{2}^{2}(\theta,\phi), \label{TN metric}
    \end{align}
    %\textcolor{red}{ \begin{align}
     %    ds^{2}=\hspace{1mm}&-f(z)\left[dt^2+4n^2(\cos\theta+\beta)^2d\phi^2 -4n(\cos\theta+\beta)d\phi dt\right]+\frac{L^4}{z^4f(z)}dz^{2}+\nonumber\\
      %   &\frac{(L^4+n^{2}z^2)}{z^2}d\Omega_{2}^{2}(\theta,\phi), \label{TN metric}
    %\end{align}}
   where we define
\begin{equation}
    f(z)=\frac{\left(z_h-z\right) \left(L^8 z z_h+n^2 z^3 z_h^3 \left(L^2+3 n^2\right)+z_h^2 \left(L^8+L^6 z^2+6 L^4 n^2 z^2\right)+L^8 z^2\right)}{L^2 z^2 z_h^3 \left(L^4+n^2 z^2\right)}.
\end{equation}
Here, \(z_h\) is the location of the horizon in the $z$ coordinate, and the boundary of TN-AdS space-time is located at \(z=0\). Next, we re-express the Hawking temperature (\ref{htem}) associated with the TN-AdS black hole in the \(z\) coordinate
\begin{align}
    T=\frac{1}{4 \pi z_hL^4}\left[z_h^2 \left(L^2+3 n^2\right)+3L^4\right]. \label{T as zh}
\end{align}

Using (\ref{T as zh}), one can further express the horizon radius ($z_h$) in terms of the temperature ($T$) as 
\begin{align}
   z_h=\frac{3 L^2}{2 \pi  L^2 T+\sqrt{4 \pi ^2 L^4 T^2-3( L^2+3 n^2)}}.\label{imgzh} 
\end{align}

Notice that the expression (\ref{imgzh}) becomes imaginary if the temperature of the TN-AdS black hole, $T<\sqrt{3} \sqrt{L^2+3 n^2}/2 \pi L^2$. To obtain a real value, we introduce a minimum limit of temperature such that
\begin{align}
    z_h=\frac{3 }{2 \pi   T \pm 2 \pi \sqrt{ T^2-T_{min}^2}}\hspace{1mm},\hspace{2mm}  T_{min}=\frac{\sqrt{3} \sqrt{L^2+3 n^2}}{2 \pi  L^2},\label{miniT}
\end{align}
where $T_{min}$ is the minimum temperature of the TN-AdS black hole.  If the temperature falls below this minimum threshold ($T<T_{min}$), the black hole solution does not exist. Instead, we have only a thermal gas configuration \cite{Abdusattar:2023fdm}-\cite{Li:2008xw}. Notice that $T_{min}$ (\ref{miniT}) gets modified due to the presence of the NUT parameter $(n)$. In other words, for large $n$, the minimum temperature of the TN-AdS black hole also becomes large.
  
In the forthcoming Sections, we will explore the effects of the NUT parameter ($n$) on the boundary observables dual to the TN-AdS black hole \eqref{TN metric}. In particular, we will investigate how the NUT parameter ($n$) affects the conductivity associated with the massless charge carriers at the boundary. 

Following the spirit of the work \cite{Karch:2007pd}, we probe the TN-AdS black hole \eqref{TN metric} with a D-brane that introduces massless charge carriers at the boundary. Next, we apply an external electric field that drifts these massless charge carriers. Under the external force, these charges move and generate a current in the direction of the electric field. Finally, one can compute the conductivity of these charge carriers from the dynamics of the D-brane. 

The dynamics of the D-brane is described by the Dirac-Born-Infeld (DBI) action \cite{Karch:2007pd}. The DBI action for an $n$-dimensional manifold can be expressed as 
\begin{equation}\label{dbiactgen}
    S_{DBI}=-T_p\int d^n\xi \sqrt{-det(g_{ab}+2\pi \alpha'F_{ab})},
\end{equation}
where  \(\xi \) are the world volume coordinates, \(g_{ab}\) is the induced world volume metric and \(T_P\) is the D-brane tension. Furthermore, here \(F_{ab}=\partial_aA_b-\partial_bA_a\) is the \(U(1)\) gauge field strength tensor on the world volume of the D-brane.

As emphasized above, the D-brane introduces \(U(1)\) charge carriers at the boundary. These charge carriers give rise to various transport phenomena such as DC conductivity \cite{Karch:2007pd},  AC conductivity \cite{Hartnoll:2009ns}, and Hall conductivity \cite{Kokado:2006ti}. In the present work, we will explore the DC conductivity in detail and leave the other phenomenon for future investigation.

\section{Holographic DC conductivity \label{DC conductivity}}  
In this Section, we study the holographic DC conductivity associated with the TN-AdS black hole \cite{Karch:2007pd} and investigate the effects of the NUT parameter ($n$) on it. To begin with, we write down the DBI action (\ref{dbiactgen}) in four dimensions as follows
\begin{equation}
    S_{DBI}=-T_p\int dt d\phi dz d \theta \sqrt{- det(g_{ab}+2\pi \alpha'F_{ab})} \label{S indp of t,phi}.
\end{equation} 

Next, we embed the D-brane trivially, i.e. \(g_{ab}=g_{\mu\nu}\), where $g_{\mu\nu}$ is the TN-AdS black hole metric (\ref{TN metric}) and consider the following ansatz for $U(1)$ gauge fields, 
\begin{align}\label{ansatzu1g}
     A_{t}=A_t(z),  A_\phi=A_\phi(z,t)=-Et+H(z),
\end{align}
where $E$ is the external electric field in the \(\phi \) direction. 

Using (\ref{TN metric}) and (\ref{ansatzu1g}) in (\ref{S indp of t,phi}), we obtain
\begin{align} S_{DBI}=-\int dz d \theta \sqrt{g_{\theta \theta}}\sqrt{|g_{tt}|g_{zz}g_{\phi \phi}+|g_{tt}|H^{'2}-g_{\phi \phi}A_t^{'2}-2|g_{t \phi}|A_t'H'+g_{zz}(|g_{t \phi}|^2-E^2) }, \label{DBI indp of A,H}
\end{align}
where $'$ denotes the derivative with respect to the variable $z$.\footnote{Here, we absorb the overall constant, $2\pi t_0 T_P$ into the definition of the DBI action.}

Notice that the action (\ref{DBI indp of A,H}) depends on functions $A_t(z)$ and $H(z)$ through their derivatives with respect to $z$. This implies that we have two conserved charges ($c$ and $d$) associated with $A_t(z)$ and $H(z)$, which are given below\footnote{Here, we use the convention $S_{DBI}=\int d^n\xi L_{DBI}$.}
\begin{align}
c=\frac{\partial L_{DBI}}{\partial A_t'}=\frac{\sqrt{g_{\theta \theta}}({A_t}'g_{\phi \phi}+|g_{t \phi}|{H}')}{\sqrt{|g_{tt}|g_{zz}g_{\phi \phi}+|g_{tt}|{H}^{'2}-g_{\phi \phi}{A_t}^{'2}-2|g_{t \phi}|{A_t}'{H}'+g_{zz}(|g_{t \phi}|^2-E^2) }},
\end{align}
and
\begin{align}
d=\frac{\partial L_{DBI}}{\partial H'}  =\frac{\sqrt{g_{\theta \theta}}(A_t'g_{t\phi}-|g_{tt}|H')}{\sqrt{|g_{tt}|g_{zz}g_{\phi \phi}+|g_{tt}|H^{'2}-g_{\phi \phi}A_t^{'2}-2|g_{t \phi}|A_t'H'+g_{zz}(|g_{t \phi}|^2-E^2) }}.  
\end{align}

After a straightforward calculation, one can eliminate the fields $A_t'$ and $H'$ and express them in terms of the conserved charges ($c$ and $d$) and TN-AdS metric components as
\begin{align}
    A_t^{'2}=\frac{g_{zz}(c|g_{tt}|+d|g_{t \phi}|)^2(|g_{tt}|g_{\phi \phi}-E^2+g_{t \phi}^2)}{(|g_{tt}|g_{\phi \phi}+g_{t \phi}^2)(c^2g_{tt}-d^2g_{\phi \phi}+g_{\theta \theta}(|g_{tt}|g_{\phi \phi}+g_{t \phi}^2)+2cd|g_{t\phi}|)},\label{A'}
\end{align}
and
\begin{align}
    H^{'2}=\frac{g_{zz}(c|g_{t \phi}|-dg_{\phi \phi})^2(|g_{tt}|g_{\phi \phi}-E^2+g_{t \phi}^2)}{(|g_{tt}|g_{\phi \phi}+g_{t \phi}^2)(c^2g_{tt}-d^2g_{\phi \phi}+g_{\theta \theta}(|g_{tt}|g_{\phi \phi}+g_{t \phi}^2)+2cd|g_{t\phi}|)}. \label{H'}
\end{align}

Using \eqref{DBI indp of A,H} and \eqref{A'}-\eqref{H'}, one finds the on-shell DBI Lagrangian as
\begin{equation}
    L_{DBI}^{on-shell}=-\frac{(L^4+n^{2}z^2)^{3/2}L^2}{z^5} \sin\theta\sqrt{\frac{|g_{tt}|g_{\phi \phi}-E^2+|g_{t \phi}|^2}{c^2 |g_{tt}|-d^2g_{\phi \phi}+|g_{tt}|g_{\phi \phi}g_{\theta \theta}+|g_{t\phi}|^2g_{\theta \theta}+2dc|g_{t \phi}|}}, \label{onshell action}
\end{equation}
where we use the metric components given in (\ref{TN metric}).

Notice that the numerator and denominator under the square root of (\ref{onshell action}) are both negative at the horizon ($z=z_h$) and positive at the boundary of space-time \((z =0)\). Therefore, for a real Lagrangian, the numerator and the denominator must change their sign at some intermediate point, say at $z=z_*$. We can obtain this intermediate point by demanding that both the numerator and the denominator vanish at $z=z_*$, which yields the following set of equations
\begin{align}
   &\hspace{1mm} |g_{tt}(z_*)|g_{\phi \phi}(z_*)+g_{t \phi}^2(z_*)-E^2=0, \label{find z_* constrent}\\
      &\hspace{1mm} c^2 |g_{tt}(z_*)|-d^2g_{\phi \phi}(z_*)+|g_{tt}(z_*)|g_{\phi \phi}(z_*)g_{\theta \theta}(z_*) +|g_{t\phi}(z_*)|^2g_{\theta \theta}(z_*)+2dc|g_{t \phi}(z_*)|=0. \label{2nd constent to find d^2}
\end{align}

Using \eqref{2nd constent to find d^2}, one can express the conserved charge \(d\) in terms of other conserved charge $c$ and the metric component as
\begin{align}
d^2=&\hspace{1mm} g_{\theta \theta}|g_{tt}|+c^2 \frac{|g_{tt}|}{g_{\phi \phi}}+2c^2\Omega^2\pm2c\Omega\sqrt{g_{\theta \theta}|g_{tt}|+c^2 \frac{|g_{tt}|}{g_{\phi \phi}}+c^2\Omega^2+\Omega^2 g_{\phi \phi}g_{\theta\theta}}\nonumber\\
&\hspace{1mm}+\Omega^2 g_{\phi \phi}g_{\theta\theta}, \label{d^2 without comp form}
\end{align}
where $\Omega = \frac{|g_{t \phi }|}{g_{\phi \phi}}$ is the angular velocity of the rotation of the space-time due to the frame dragging  \cite{Durka:2019ajz}.

It is important to notice that the above equation (\ref{d^2 without comp form}) is valid only at $z=z_*$. Therefore, we next aim to determine this intermediate point ($z_*$). In order to obtain the $z_*$, we plug (\ref{TN metric}) in (\ref{find z_* constrent}), which yields  
\begin{align}
    & \left(z_h-z_*\right) \sin^2\theta \left[L^8 z_* z_h+n^2 z_*^3 z_h^3 \left(L^2+3 n^2\right)+z_h^2 \left(L^8+L^6 z_*^2+6 L^4 n^2 z_*^2\right)+L^8 z_*^2\right]\nonumber\\
    &-E^2 L^2 z_*^4 z_h^3=0. \label{exact eqn of z_*}
\end{align}

Notice that the above equation (\ref{exact eqn of z_*}) is technically challenging to solve exactly in the electric field $E$. However, we can solve it perturbatively, treating the external electric field $E$ as an expansion parameter, i.e., $E<<1$.

One can systematically expand $z_*$ as follows
\begin{align}\label{expansion}
    z_*=z_*^{(0)}-E^2 z_*^{(1)}-E^4z_*^{(2)}\hspace{2mm},\hspace{1mm} E^2<<1.
\end{align}
Here $z_*^{(0)}$ is the solution of (\ref{exact eqn of z_*}) for the vanishing external electric field $E$, $z_*^{(1)}$ is the leading order correction and $z_*^{(2)}$ is the next to leading order (NLO) correction in the presence of $E$.

Using (\ref{expansion}) in (\ref{exact eqn of z_*}), we obtain $z_*$ upto NLO,
\begin{align}
    z_*=&\hspace{1mm}z_h-E^2\Big[\left( L^2 z_h^5 \csc ^2\theta\right)\left(L^6 z_h^2+6 L^4 n^2 z_h^2+L^2 n^2 z_h^4+3 n^4 z_h^4+3 L^8\right)^{-1}\Big]+\nonumber\\
    &\hspace{1mm}E^4\Big[\left( L^4 z_h^9 \csc ^4\theta \left(n^2 z_h^4 \left(L^2+3 n^2\right)+2 z_h^2 \left(L^6+6 L^4 n^2\right)+9 L^8\right)\right)\times\nonumber\\
    &\hspace{1mm}\left(n^2 z_h^4 \left(L^2+3 n^2\right)+z_h^2 \left(L^6+6 L^4 n^2\right)+3 L^8\right){}^{-3}\Big]\label{z_* upto E^2}.
\end{align}

Next, we substitute (\ref{z_* upto E^2}) into (\ref{d^2 without comp form}).  After identifying the conserved charge $d= <J_\phi>\sin\th$ with the current density and $c=<J_t>\sin\th$ as the charge density \cite{Karch:2007pd}, we obtain %\ref{Ab}\footnote{See Appendix \ref{Ab}, for detailed discussion. }
\begin{align}
   <J_\phi>^2=&\hspace{1mm}\frac{J_t^2}{\sin^2\th}\left[ \frac{E^2 z_h^4 \csc ^2\theta }{\left(n^2 z_h^2+L^4\right){}^2}-\frac{4 E^4 L^6 z_h^8 \csc ^4\theta  }{\left(n^2 z_h^2+L^4\right){}^4 \left(z_h^2 \left(L^2+3 n^2\right)+3 L^4\right)}+3 \Omega ^2 \sin ^2\theta \right] \nonumber \\ 
   &\hspace{1mm}+ E^2 \csc ^4\theta+\frac{\Omega ^2 \left(n^2 z_h^2+L^4\right){}^2}{z_h^4}+\frac{2 E \Omega  J_t \sqrt{z_h^4 J_t^2+\left(n^2 z_h^2+L^4\right){}^2}}{\sin^2\th(n^2 z_h^2+L^4)},\label{condc}
\end{align}
where we identify
\begin{align}
    | \Omega|=\frac{E^2}{\sin^4\theta}\frac{2n z_h^4|\left(\cos \theta+\beta\right)|}{\left(n^2 z_h^2+L^4\right){}^2} +O(E^4).\label{omega at z*}
\end{align}
Notice that we truncate $\Omega$ (\ref{omega at z*}) upto $E^2$ because it appears as a square term or is coupled with the external electric field in (\ref{condc}). Therefore, $O(E^4)$ corrections in $\Omega$ will contribute to (\ref{condc}) beyond the quartic level.%\footnote{we have constraint on \(\theta\) at \(z_*\) for \(\beta=1\) ;\(-2\sin^{-1}\sqrt{\frac{z_h E n}{L^4 +z_h^2n^2}}<\theta\leq\pi\),\\for \(\beta=0\) ;\(\sin^{-1}\sqrt{\frac{2z_h^2En}{L^4+n^2z_h^2}}<\theta<-\sin^{-1}\sqrt{\frac{2z_h^2En}{L^4+n^2z_h^2}}\)\\ for \(\beta=-1\);\(0 \leq \theta < 2\cos^{-1}\sqrt{\frac{z_h E n}{L^4 +z_h^2n^2}}\)}

After comparing the above expression (\ref{condc}) with $<J_{\phi}>=\sigma E$, one finds the DC conductivity ($\sigma$) in the dual QFT as
\begin{align}
    \sigma=&\hspace{1mm} \sqrt{\s_{U(1)}^2+\s_{thermal}^2}\hspace{1mm},\label{FC}
\end{align}
where we divide the contribution to the conductivity into following two parts\footnote{For the rest of our analysis, we shall rescale conductivities by the overall factor of $\sin^4\th$.}
\begin{align}
    \s^2_{U(1)} = &\hspace{1mm}J_t\Bigg[\frac{z_h^4J_t }{\left(n^2 z_h^2+L^4\right){}^2}\left(1 - \frac{4e^2  z_h^6}{\sin^2\theta}\frac{L^6}{\left(n^2 z_h^2+L^4\right){}^2 \left(\left(3 n^2+L^2\right) z_h^2+3L^4\right)}\right)\nonumber\\
    &\hspace{1mm}+\frac{3J_te^2z_h^2}{\sin^2\theta}\tilde\Omega^2+\frac{2ez_h}{\sin\theta}\tilde\Omega\sqrt{1+\frac{J_t^2 z_h^4 }{\left(n^2 z_h^2+L^4\right){}^2}}\Bigg],\label{sig1}\\
  \sigma_{thermal}^2= &\hspace{1mm}  1+\frac{e^2}{\sin^2\theta}{\frac{(n^2z_h^2+L^4)^2\tilde\Omega^2}{ z_h^2}}.\label{sig2}
\end{align}
Here we introduce the dimensionless quantity $e=E/z_h$ and $\tilde\Omega=\Omega\sin^3\theta/E^2$.

The above expression for conductivity (\ref{FC}) is the key finding of our analysis. It is worth noting that the TN-AdS conductivity (\ref{FC}) bears similarities to the conductivity computed for the AdS\(_5\) planar black hole, as discussed by the authors in \cite{Karch:2007pd}. They proposed that conductivity is due to two types of charge carriers. The first type is explicitly introduced by U(1) charge carriers, being proportional to $< J_t >$, while the second type comes from thermally produced charge pairs.

The conductivity (\ref{FC}) associated with the TN-AdS black hole also has a similar interpretation. The first term, denoted as ($\s_{U(1)}$) (\ref{sig1}), is due to the added charge carriers and is proportional to $< J_t >$. On the other hand, the second term (\ref{sig2}) is the thermal contribution to the conductivity. Notably, both the conductivities ($\s_{U(1)}$ and $\s_{thermal}$) are significantly affected due to the presence of the NUT parameter $(n)$ and the frame dragging  ($\Omega$). We discuss these aspects in more detail in the following Section.

\section{NUT parameter and frame dragging effects}\label{nuteffectsec}
 In this Section, we examine the effects of the NUT parameter $(n)$ and the frame dragging $(\Om)$ on the conductivity (\ref{FC}) associated with the TN-AdS black hole. In particular, we study $\s_{U(1)}$ (\ref{sig1}) as well as $\s_{thermal}$ (\ref{sig2}) for various values of $\beta = 0,\pm1$, and we explore their dependency on the temperature ($T$) both near and far away from the Misner string. 
 
 To begin with, we first write the angular rotational velocity \(\tilde\Omega\) (\ref{omega at z*}) for $\b=0,\pm1$ as a function of temperature $T$ (\ref{miniT}), which is given below\footnote{We set the AdS length scale $L=1$ for the rest of the analysis.}
 \begin{align}
     \big |\tilde\Omega\big|_{\b=1}=&\hspace{1mm}\frac{162 n }{\left(4 \pi ^2  \left(2 T \tilde{T}-T_{min }^2+2 T^2\right)+9 n^2\right)^2}\cot \left(\frac{\theta }{2}\right),\label{av1}\\
      \big| \tilde\Omega\big|_{\b=0}=&\hspace{1mm}\frac{162  n }{\left(4 \pi ^2  \left(2 T \tilde{T}-T_{min }^2+2 T^2\right)+9 n^2\right)^2}\cot (\theta ),\label{angb1}\\
   \big| \tilde\Omega\big|_{\b=-1}=&\hspace{1mm}\frac{162  n }{\left(4 \pi ^2  \left(2 T \tilde{T}-T_{min }^2+2 T^2\right)+9 n^2\right)^2}\tan \left(\frac{\theta }{2}\right),\label{angb2}
 \end{align}
where we define $\tilde{T}=\sqrt{T^2-T_{min }^2}$.

Next, we substitute the expression for $z_h$ (\ref{miniT}) into $(\ref{omega at z*})$, as well as both $\s_{U(1)}$ (\ref{sig1}) and $\s_{thermal}$ (\ref{sig2}). This yields the following results\footnote{Notice that the effects of frame dragging $(\tilde\Omega)$ are included in angle $\theta$, which we elaborate next. Furthermore, both the conductivities ($\s_{U(1)}$ and $\s_{thermal}$) are finite at the minimum temperature ($T_{min}$).}:

%Using, (\ref{av1})-(\ref{angb2}), one can express $\s_{U(1)}$ (\ref{sig1}) and $\s_{thermal}$ (\ref{sig2}) as a function of temperature (\ref{miniT}) as follows
\begin{align}
&\s_{U(1)}= \hspace{1mm}\frac{9  J_t}{\left(4 \pi ^2  \left(2 T \tilde{T}-T_{min }^2+2 T^2\right)+9 n^2\right)}\Bigg[\frac{972 E^2 n^2 \csc ^4\theta  (\beta +\cos \theta )^2}{\left(4 \pi ^2  \left(2 T \left(\tilde{T}+T\right)-T_{min }^2\right)+9 n^2\right)^2}
\nonumber\\
&-\frac{432 \pi ^2 E^2 \csc ^2\theta \left(\tilde{T}+T\right){}^2}{\left(8 \pi ^2 T \left(\tilde{T}+T\right)-4 \pi ^2 T_{min }^2+9 n^2\right)^2 \left(8 \pi ^2 T \left(\tilde{T}+T\right)-4 \pi ^2 T_{min }^2+9 n^2+3\right)}\nonumber\\
& +\frac{4 E n \csc ^2\theta  (\beta +\cos \theta ) \left(81 J_t^2+\left(4 \pi ^2  \left(\tilde{T}+T\right)^2+9 n^2\right)^2\right)^{1/2}}{J_t \left(8 \pi ^2  T \left(\tilde{T}+T\right)-4 \pi ^2  T_{min }^2+9 n^2\right)}\hspace{2mm}+1\Bigg]^{1/2}\label{UTT},\\
&\s_{thermal}= \hspace{1mm}\sqrt{1+\frac{324 E^2 n^2 \csc ^4\theta  (\beta +\cos \theta )^2}{\left(4 \pi ^2  \left(2 T \left(\tilde{T}+T\right)-T_{min }^2\right)+9 n^2\right)^2}}\label{ETT}.
\end{align}
In the following, we separately examine the conductivities (\ref{UTT})-(\ref{ETT}) in both low and high temperature regimes.

 \subsection{Low temperature regime}
We define the low temperature regime in the limit where the temperature is close to the minimum temperature ($T\sim T_{min}$) (\ref{miniT}). We first consider the case where $\b=1$. As mentioned in Section \ref{ri}, when $\b=1$, the Misner string is located at $\th=0$.
%Next, we examine the effects of frame dragging (\ref{omega at z*}) on the conductivity (\ref{FC}). It is important to notice that these effects become relevant only at NLO. However, their impact can be significant depending on the position of the Misner string. Below, we discuss the frame dragging effects on both $\s_{U(1)}$ and $\s_{thermal}$. 

It is important to notice that the novel frame dragging effects become relevant only at NLO. However, their impact can be significant depending on the position of the Misner string. For instance, when we are far from the Misner string ($\theta\sim\pi$), the value of $\tilde{\Omega}$ (\ref{av1}) becomes negligibly small. On the other hand, as we approach the Misner string ($\theta \sim 0$), frame dragging effects start to dominate. Therefore, we will examine the $\s_{U(1)}$ (\ref{UTT}) as well as $\s_{thermal}$ (\ref{ETT}) near the Misner ($nM$) string, and far away ($fM$) from it.

%Before proceeding further, we first note the angular rotational velocity \(\tilde\Omega\) (\ref{omega at z*}) for $\b=1$, which is given below
%\begin{align}
 %   |\tilde\Omega|=\frac{162 n }{\left(4 \pi ^2  \left(2 T \tilde{T}-T_{min }^2+2 T^2\right)+9 n^2\right)^2}\cot \left(\frac{\theta }{2}\right)\label{av1}
%\end{align}

In the low temperature regime, both $\s_{U(1)}$ (\ref{UTT}) and $\s_{thermal}$ (\ref{ETT}) can be expressed as follows
\begin{align}
   & \sigma_{U(1)}\Big |_{T\sim T_{min}}^{fM}\approx\frac{9J_t}{4\pi^2T_{min}^2 }+O\left(\sqrt{T-T_{min }}\right),\label{su1aa}\\
   &\sigma_{U(1)}\Big |_{T\sim T_{min}}^{nM} \approx\frac{81 \sqrt{3} E nJ_t}{4 \pi ^4 \theta^2T_{min }^4} +O\left(\sqrt{T-T_{min }}\right),\label{su1}\\
    &\sigma_{thermal}\Big |_{T\sim T_{min}}^{fM} \approx1,\sigma_{thermal} \Big |_{T\sim T_{min}}^{nM} \approx\frac{9 E n }{ \pi ^2 \theta^2 T_{min }^2}+O\left(\sqrt{T-T_{min }}\right),\label{st1}
\end{align}
where $0<\th<<\pi$ measures how close we are to the Misner string.
%\textcolor{red}{\begin{align}
 %  & \sigma_{U(1)}\Big |_{T\sim T_{min}}^{fM}\approx\frac{9J_t}{4\pi^2T_{min}^2 }\hspace{1mm},\hspace{2mm} \sigma_{U(1)}\Big |_{T\sim T_{min}}^{nM} \approx\frac{81 \sqrt{3} E nJ_t}{16 \pi ^4 T_{\min }^4} \csc ^2\left(\frac{\theta }{2}\right) +O(T^0_{min})\label{su1}\\
  %  &\sigma_{thermal}\Big |_{T\sim T_{min}}^{fM} \approx1\hspace{1mm},\hspace{2mm}\sigma_{thermal} \Big |_{T\sim T_{min}}^{nM} \approx\frac{9 E n }{4 \pi ^2  T_{\min }^2}\csc ^2\left(\frac{\theta }{2}\right)+O({T^2_{min}})\label{st1}
%\end{align}}

\begin{figure}[H]
     \centering
     \begin{subfigure}[b]{0.495\textwidth}
         \centering
         \includegraphics[width=\textwidth]{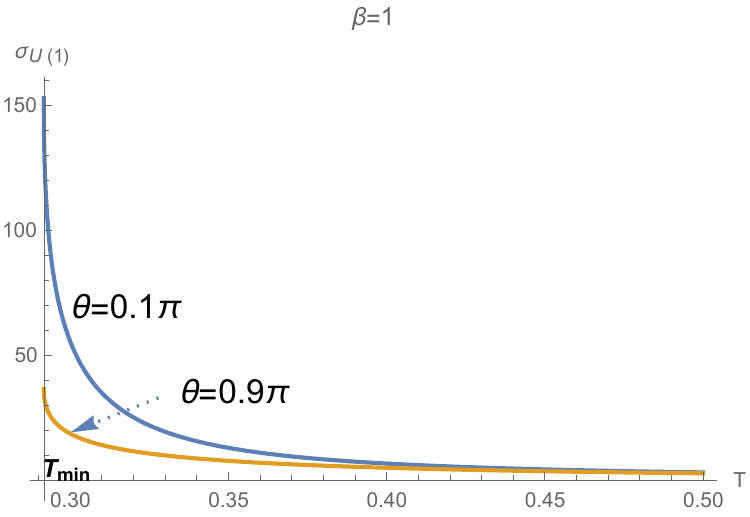}
         \caption{$\s_{U(1)}$ vs temperature}\label{figure1a}
              \end{subfigure}
     \hfill
     \begin{subfigure}[b]{0.495\textwidth}
         \centering
         \includegraphics[width=\textwidth]{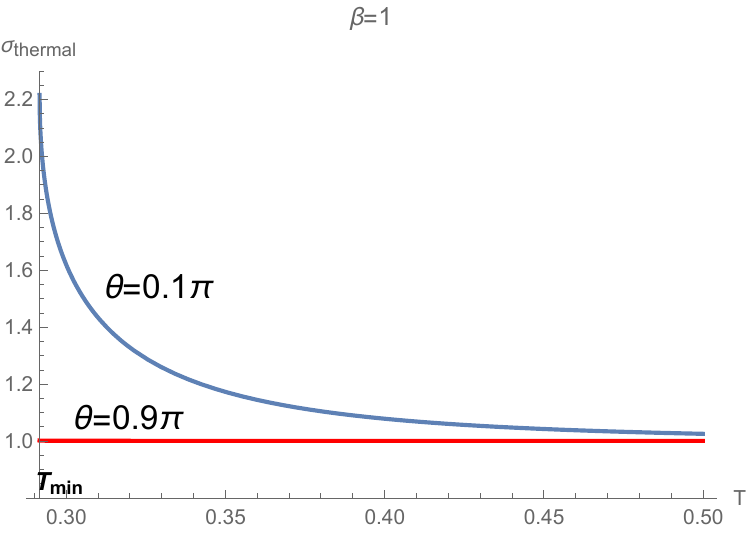}
          \caption{$\s_{thermal}$ vs temperature} \label{figure1b}
             \end{subfigure}
             \caption{Conductivity vs temperature plot for $J_t=10$, $E=0.1$ and $T_{min}=0.29$.}
        \label{figure1}
\end{figure}

Below, we plot both the conductivities (\ref{UTT}) and (\ref{ETT}) against temperature (see Figure \ref{figure1}) for the NUT parameter $n=0.2$. For points near the Misner ($nM$) string, we set $\theta=0.1\pi$, and for points away ($fM$) from the Misner string, we set $\theta=0.9\pi$.
%\newpage

Figure \ref{figure1} illustrates the temperature dependencies of both thermal conductivity ($\s_{thermal}$) and the conductivity  ($\s_{U(1)}$) due to the added U(1) charge carriers. Our analysis reveals that $\s_{U(1)}$ (\ref{su1aa})-(\ref{su1}) increases sharply as the temperature approaches the minimum threshold ($T\sim T_{min}$). However, the increase (near the Misner string) is even sharper due to the frame dragging effects (see Figure \ref{figure1a}). Physically, this stems from the fact that the U(1) charge carriers are drifted more rapidly due to the effects of frame dragging, causing an increase in the conductivity ($\s_{U(1)}$).

On a similar note, \(\sigma_{thermal}\) (\ref{st1}) also increases as $T\sim T_{min}$ and the effect is prominent close to the Misner string only, while it remains nearly constant (with temperature) at points located further away from it (see Figure \ref{figure1b}). The reason is similar to above, namely, the thermally produced charge pairs that are away from the Misner string are less drifted than those closer to the string.

The above statement can be justified better by taking the ratio of \(\sigma_{U(1)}\) (\ref{su1}) near the Misner string $(nM$) to that far away ($fM$) from it, and similarly for \(\sigma_{thermal}\) (\ref{st1}), which yields the following results upto leading order:
%\textcolor{red}{\begin{align}
 %   \sigma_{U(1)}\Big |_{T\sim T_{min}}^{\theta\sim\pi}\approx\frac{9J_t}{4\pi^2T_{min}^2\sin^2\theta }\hspace{1mm},\hspace{2mm} \sigma_{U(1)}\Big |_{T\sim T_{min}}^{\theta\sim0} \approx\frac{9  J_t \sqrt{16 \pi ^4 L^8 T_{\min }^4+243 E^2 n^2 \csc ^4\left(\frac{\theta }{2}\right)}}{16 \pi ^4 L^8 \sin^2(\theta) T_{\min }^4}\hspace{1mm}
%\end{align}$$\frac{\lim\limits_{\theta \to \epsilon}\sigma_{thermal}}{\lim\limits_{\hspace{2mm}\theta \to \pi-\epsilon}\sigma_{thermal}}\bigg|_{T \sim T_{min}}  \approx \sqrt{1+\frac{243 E^2 n^2}{\pi ^4 L^8 \epsilon ^4 T_{\min }^4}}>>1$$

%\begin{align}
 %\sigma_{thermal}\Big |_{T\sim T_{min}}^{\theta\sim\pi} \approx\frac{1}{\sin^2\theta}.\hspace{1mm},\hspace{2mm}\sigma_{thermal} \Big |_{T\sim T_{min}}^{\theta\sim0} \approx\frac{1}{\sin^2\theta}\sqrt{\left(1+\frac{81 E^2 n^2 \csc ^4\left(\frac{\theta }{2}\right)}{16 \pi ^4 L^8 T_{min}^4}\right)}\label{nn1}
%\end{align}
%$$\frac{\lim\limits_{\theta \to \epsilon}\hspace{2mm}\sigma_{thermal}}{\lim\limits_{\hspace{2mm}\theta \to \pi-\epsilon}\sigma_{thermal}}\bigg|_{T \sim T_{min}} =\sqrt{\left(1+\frac{81n^2E^2}{\pi^4L^8\epsilon^4T_{min}^4}\right)}>>1 $$
\begin{align}
    \frac{\sigma_{U(1)}\bigg|_{T \sim T_{min}}^{nM}}{\sigma_{U(1)}\bigg|_{T \sim T_{min}}^{fM}}  = \frac{9\sqrt{3} En}{\pi ^2\theta^2 T_{min }^2}>>1\hspace{1mm},\hspace{2mm}\frac{\sigma_{thermal}\bigg|_{T \sim T_{min}}^{nM}}{\sigma_{thermal}\bigg|_{T \sim T_{min}}^{fM}}=\frac{9nE}{\pi^2\theta^2T_{min}^2}>>1.\label{ratio}
\end{align}
%\textcolor{red}{\begin{align}
 %   \frac{\sigma_{U(1)}\bigg|_{T \sim T_{min}}^{nM}}{\sigma_{U(1)}\bigg|_{T \sim T_{min}}^{fM}}  = \frac{9\sqrt{3} En}{\pi ^2\epsilon^2 T_{min }^2}>>1\hspace{1mm},\hspace{2mm}\frac{\sigma_{thermal}\bigg|_{T \sim T_{min}}^{nM}}{\sigma_{thermal}\bigg|_{T \sim T_{min}}^{fM}}=\frac{9nE}{\pi^2\epsilon^2T_{min}^2}>>1\label{ratio}
%\end{align}}
%\footnote{\textcolor{blue}{For points close to the Misner string, $\theta\sim\epsilon$, while for points farther away, $\theta\sim\pi-\epsilon$.\label{fn1}}.}

To summarize, in the low temperature regime ($T\sim T_{min}$), both the conductivities $\sigma_{U(1)}$ and $\sigma_{thermal}$ are greater near the Misner string compared to the regions that are farther away from it (\ref{ratio}). Notice that the above phenomenon is sourced due to the presence of the NUT parameter ($n$), which causes frame dragging effects (\ref{av1}) in the theory.

\vspace{2mm}
$\bullet$ \textbf{For cases $\b=0$ and $\b=-1$} 

\vspace{1mm}
It is worth noting that one can also perform a similar calculations for the other two cases, i.e., $\b=0$ and $\b=-1$. In the case $\b=0$, the Misner string is located either\footnote{In the present analysis, we consider the location of the Misner string for $\beta=0$ to be at $\theta=\pi$.} at $\th=0$ or $\th=\pi$. However, for $\b=-1$, the Misner string is located only at $\th=\pi$, as shown in Figure \ref{fig:placeholder}. 

To proceed further, we express both $\sigma_{U(1)}$ (\ref{UTT}) and $\sigma_{thermal}$ (\ref{ETT}) in the regions near and far from the Misner string. In the low temperature regime ($T\sim T_{min}$), for $\beta=0$, we find
\begin{align}
   &\sigma_{U(1)}\Big |_{T\sim T_{min}}^{fM}\approx\frac{9J_t}{4\pi^2T_{min}^2 }+O\left(\sqrt{T-T_{min}}\right),\label{ex1aa}\\
  &\sigma_{U(1)}\bigg|^{nM}_{T\sim T_{min}}\approx \frac{81 \sqrt{3} E J_t n }{8 \pi ^4 (\pi-\theta)^2 T_{min }^4} +O\left(\sqrt{T-T_{min}}\right),\label{ex1}\\
  &\sigma_{thermal}\Big |_{T\sim T_{min}}^{fM} \approx1,
\sigma_{thermal}\bigg|^{nM}_{T \sim T_{min}}\approx \frac{9 E n }{2 \pi ^2 (\pi-\theta)^2 T_{min }^2}+O\left(\sqrt{T-T_{min}}\right)\label{ex2}.
\end{align}
%\textcolor{red}{\begin{align}
 %  &\sigma_{U(1)}\Big |_{T\sim T_{min}}^{fM}\approx\frac{9J_t}{4\pi^2T_{min}^2}, 
  %\sigma_{U(1)}\bigg|^{nM}_{T\sim T_{min}}\approx \frac{81 \sqrt{3} E J_t n \cot (\theta ) \csc (\theta )}{8 \pi ^4 T_{\min }^4}+O(T^0_{min}),\label{ex1}\\&\sigma_{thermal}\Big |_{T\sim T_{min}}^{fM} \approx1\hspace{1mm},\hspace{2mm}
%\sigma_{thermal}\bigg|^{nM}_{T \sim T_{min}}\approx \frac{9 E n \cot (\theta ) \csc (\theta )}{2 \pi ^2  T_{\min }^2}+O(T^2_{min})
 %\label{ex2}.
%\end{align}}
Similarly, for $\beta=-1$, we have
\begin{align}   &\sigma_{U(1)}\Big |_{T\sim T_{min}}^{fM}\approx\frac{9J_t}{4\pi^2T_{min}^2 }+O\left(\sqrt{T-T_{min}}\right),\label{ex3aa}\\&\sigma_{U(1)}\bigg|^{nM}_{T\sim T_{min}}\approx \frac{81 \sqrt{3} En J_t}{4 \pi ^4 (\pi-\th)^2 T_{min }^4}  +O\left(\sqrt{T-T_{min}}\right),\label{ex3}\\&\sigma_{thermal}\Big |_{T\sim T_{min}}^{fM} \approx1,\sigma_{thermal}\bigg|^{nM}_{T \sim T_{min}}\approx\frac{9 E n}{\pi ^2 (\pi-\theta)^2 T_{min }^2}+O\left(\sqrt{T-T_{min}}\right),\label{ex4}
\end{align}
where $0<<\th<\pi$ measures how close we are to the Misner string.
%\textcolor{red}{\begin{align}   &\sigma_{U(1)}\Big |_{T\sim T_{min}}^{fM}\approx\frac{9J_t}{4\pi^2T_{min}^2 }, \sigma_{U(1)}\bigg|^{nM}_{T\sim T_{min}}\approx \frac{81 \sqrt{3} En J_t}{16 \pi ^4  T_{\min }^4} \sec ^2\left(\frac{\theta }{2}\right)+O(T^0_{min})\label{ex3}\\&\sigma_{thermal}\Big |_{T\sim T_{min}}^{fM} \approx1\hspace{1mm},\hspace{2mm}\sigma_{thermal}\bigg|^{nM}_{T \sim T_{min}}\approx\frac{9 E n}{4 \pi ^2  T_{\min }^2} \sec ^2\left(\frac{\theta }{2}\right)+O(T_{min}^2)\label{ex4}
%\end{align}}

\begin{figure}[htp]
     \centering
     \begin{subfigure}[b]{0.495\textwidth}
         \centering
         \includegraphics[width=\textwidth]{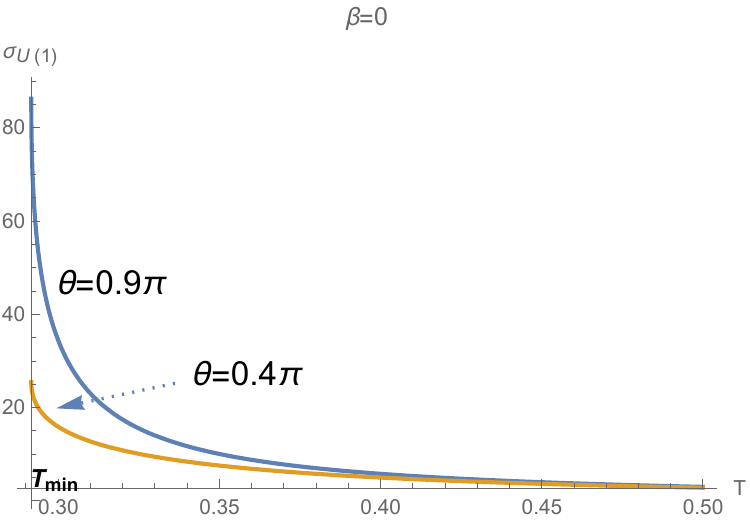}
         \caption{$\s_{U(1)}$ vs temperature}\label{0sub1}
              \end{subfigure}
     \hfill
     \begin{subfigure}[b]{0.495\textwidth}
         \centering
         \includegraphics[width=\textwidth]{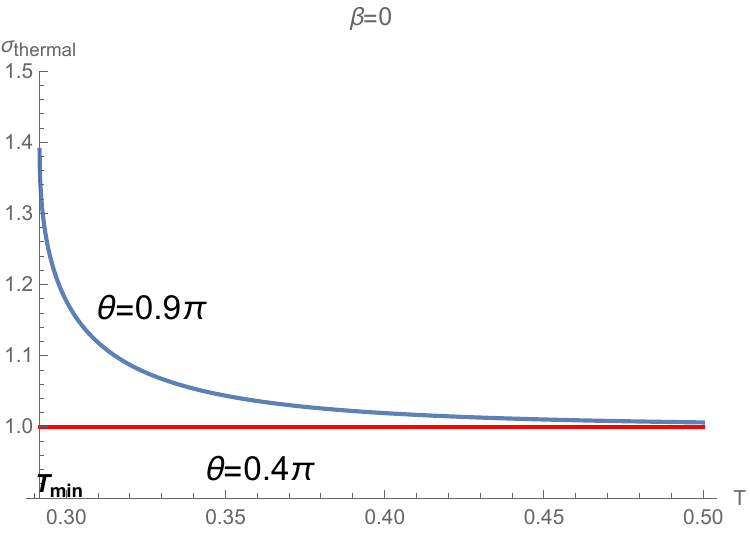}
         \caption{$\s_{thermal}$ vs temperature}\label{0sub2}
             \end{subfigure}
             \caption{Conductivity vs temperature plot for $\b=0$, $E=0.1$ and $J_t=10$}
        \label{figure2}
\end{figure}

\begin{figure}[htp]
     \centering
     \begin{subfigure}[b]{0.495\textwidth}
         \centering
         \includegraphics[width=\textwidth]{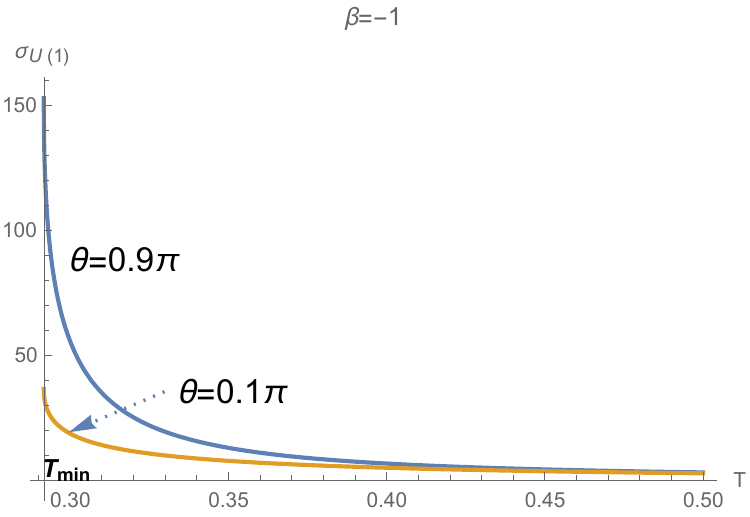}
         \caption{$\s_{U(1)}$ vs temperature}\label{m1sub1}
              \end{subfigure}
     \hfill
     \begin{subfigure}[b]{0.495\textwidth}
         \centering
         \includegraphics[width=\textwidth]{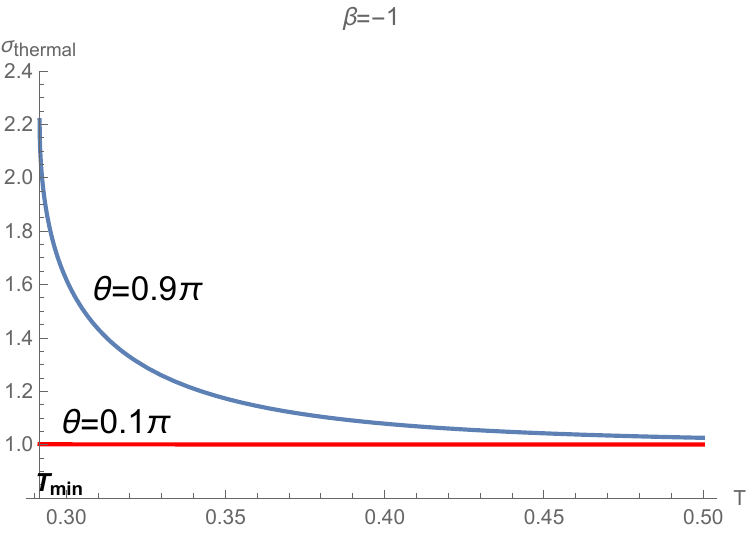}
         \caption{$\s_{thermal}$ vs temperature}\label{m1sub2}
             \end{subfigure}
             \caption{Conductivity vs temperature plot for $\b=-1$, $E=0.1$ and $J_t=10$}
        \label{figure3}
\end{figure}

Below, we plot the $\s_{U(1)}$ (\ref{UTT}) and $\s_{thermal}$ (\ref{ETT}) as a function of temperature ($T$) for both cases, $\b=0$ and $\b=-1$, as illustrated in Figures \ref{figure2} and \ref{figure3}, respectively. In both cases, we set \(n=0.2\), which yields a minimum temperature \(T_{\text{min}}=0.29\) (\ref{miniT}). For both cases, \(\beta=0\) and \(\beta=-1\), we set \(\theta=0.9\pi\) near the Misner ($nM$) string. For points further away, we set \(\theta=0.4\pi\) for $\beta=0$ and \(\theta=0.1\pi\) for $\beta=-1$. 

%\newpage
Like before, we find that the $\sigma_{U(1)}$ (\ref{ex1aa})-(\ref{ex1}), (\ref{ex3aa})-(\ref{ex3}) increases as $T\sim T_{min}$. However, the rise near the Misner string is sharper due to the frame dragging as explained earlier (see Figures \ref{0sub1} and \ref{m1sub1}). On a similar note, $\s_{thermal}$ (\ref{ex2}), (\ref{ex4}) also increases significantly near the Misner string ($nM$), while it remains almost constant for points located far from the Misner string ($fM$) (see Figures  \ref{0sub2} and \ref{m1sub2}). This behavior is also identical to what has been observed for case $\beta = 1$.

Finally, to have a fair comparison, we take the ratio of conductivities (of the same type) (\ref{ex1})-(\ref{ex4}) near ($nM$) and far ($fM$) from the Misner string. For $\b=0$, this can be expressed upto leading order as follows
\begin{align}
     \frac{\sigma_{U(1)}\bigg|_{T \sim T_{min}}^{nM}}{\sigma_{U(1)}\bigg|_{T \sim T_{min}}^{fM}}  = \frac{9\sqrt{3} E n}{2 \pi ^2(\pi-\theta)^2 T_{min }^2}>>1\hspace{1mm},\hspace{2mm}\frac{\sigma_{thermal}\bigg|_{T \sim T_{min}}^{nM}}{\sigma_{thermal}\bigg|_{T \sim T_{min}}^{fM}}= \frac{9 E n }{2 \pi ^2 (\pi-\theta)^2T_{min}^2}>>1. \label{ratio1}
\end{align}
%\textcolor{red}{\begin{align}
 %    \frac{\sigma_{U(1)}\bigg|_{T \sim T_{min}}^{nM}}{\sigma_{U(1)}\bigg|_{T \sim T_{min}}^{fM}}  = \frac{9\sqrt{3} E n}{2 \pi ^2 \epsilon ^2 T_{\min }^2}>>1\hspace{1mm},\hspace{2mm}\frac{\sigma_{thermal}\bigg|_{T \sim T_{min}}^{nM}}{\sigma_{thermal}\bigg|_{T \sim T_{min}}^{fM}}=\frac{9 E n }{2 \pi ^2 \epsilon ^2T_{min}^2}>>1 \label{ratio1}
%\end{align}}
Similarly for $\b=-1$, we find
\begin{align}
     \frac{\sigma_{U(1)}\bigg|_{T \sim T_{min}}^{nM}}{\sigma_{U(1)}\bigg|_{T \sim T_{min}}^{fM}}  =\frac{9 \sqrt{3} E n}{\pi ^2(\pi-\theta) ^2 T^2_{min }}>>1\hspace{1mm},\hspace{2mm}\frac{\sigma_{thermal}\bigg|_{T \sim T_{min}}^{nM}}{\sigma_{thermal}\bigg|_{T \sim T_{min}}^{fM}}=\frac{9nE}{\pi^2(\pi-\theta)^2T_{min}^2}>>1.\label{ratio2}
\end{align}
%\textcolor{red}{\begin{align}
 %    \frac{\sigma_{U(1)}\bigg|_{T \sim T_{min}}^{nM}}{\sigma_{U(1)}\bigg|_{T \sim T_{min}}^{fM}}  =\frac{9 \sqrt{3} E n}{\pi ^2\epsilon ^2 T^2{}_{\min }}>>1\hspace{1mm},\hspace{2mm}\frac{\sigma_{thermal}\bigg|_{T \sim T_{min}}^{nM}}{\sigma_{thermal}\bigg|_{T \sim T_{min}}^{fM}}=\frac{9nE}{\pi^2\epsilon^2T_{min}^2}>>1,\label{ratio2}
%\end{align}}
%\footnote{\textcolor{blue}{For \(\beta = 0\), we have  either $\theta\sim\pi-\epsilon$ or $\theta\sim\epsilon$ near the Misner string, and \(\th \sim \frac{\pi}{2} - \epsilon\) when farther away from it. On the other hand, for the case \(\beta = -1\), near the Misner string, we have $\theta\sim\pi-\epsilon$  and \(\th \sim \epsilon\) for farther points.\label{fn2}}.}

%Finally, one can ignore the thermal contribution in the resistivity and find that the entire contribution comes from the explicitly introduced charged carrier. To be precise, when we are far from the Misner string, the resistivity behaves as $\rho_{U(1)}\sim T^{2}$. In contrast, near the Misner string, it increases as $\rho_{U(1)}\sim T^{4}$. This behavior is consistent for both cases, $\b=0$ and $\b=-1$. 
%\begin{table}[]
 %   \centering
%\renewcommand{\arraystretch}{1.7}
%\begin{tabular}{|c|c|c|c|c|}
 %\hline
 %$\beta$ &    $\sigma_{U(1)}/\sigma_{thermal}\Big|_{\text{nM}}^{\text{LT}}$  & $\sigma_{U(1)}/\sigma_{thermal}\Big|_{\text{fM}}^{\text{LT}}$ & $\rho\bigg|^{LT}_{nm} \approx\rho_{U(1)}\Big|_{\text{nM}}^{\text{LT}}$ & $\rho\bigg|^{LT}_{nm} \approx\rho_{U(1)}\Big|_{\text{fM}}^{\text{LT}}$
% \\ 
 %\hline
 %$1$ &$\frac{9J_t\sqrt{3}}{4\pi^2L^4T^2}>>1$
  % & $\frac{9J_t}{4\pi^2L^4 T^2}>>1$& $\frac{16\pi^4T^4}{81\sqrt{3}J_t}$ & $\frac{4\pi^2T^2}{9J_t}$\\
  %\hline
 %$0$ & $\frac{9J_t\sqrt{3}}{4\pi^2L^4T^2}>>1$&  $ \frac{9J_t}{4\pi^2 L^4T^2}>>1$&$\frac{8\pi^4T^4}{81\sqrt{3}J_t}$& $\frac{4\pi^2T^2}{9J_t}$ \\
 %\hline
%\end{tabular}
 %\caption{Conductivity and resistivity for various values for $\beta$ in the low temperature (LT) regime. Here, nM refers to near the Misner string, while fM refers to far from the Misner string.}
    %\label{tablelt}
%\end{table}

To summarize, in all three cases $\b=0,\pm1$, we find that the thermal conductivity and the U(1) conductivity (arising due to the added U(1) charge carriers of D-brane) dominate near the Misner string, as a result of the novel frame dragging that drifts the charge carriers near the Misner string (\ref{av1})-(\ref{angb2}). On the other hand, far from the  Misner string ($fM$), the frame dragging is less, and hence the corresponding conductivities are also less. 

Finally, as a crucial comparison between two different types of conductivities (at low temperatures and upto leading order), one should notice that $\s_{U(1)}$ is always greater than $\s_{thermal}$ for both near and far from the Misner string (\ref{su1})-(\ref{ex4}). This can be seen by taking the ratio 
\begin{align}
  \frac{\s_{U(1)}^{nM}}{\s_{thermal}^{nM}}\Bigg|_{\b=0,\pm1} = \frac{9\sqrt{3}J_t}{4\pi^2T_{min}^2 }>>1\hspace{1mm},\hspace{2mm}\frac{\s_{U(1)}^{fM}}{\s_{thermal}^{fM}}\Bigg|_{\b=0,\pm1}=\frac{9J_t}{4\pi^2T_{min}^2 }>>1.\label{dcratio}
\end{align}
The above result implies that, close to the threshold temperature ($T\sim T_{min}$), the electrically charged carriers are significant in number and therefore play a dominant role as compared to the thermally produced charge pairs \cite{Karch:2007pd}.

Below, in Table \ref{tableltn}, we summarize all the key findings at low temperature ($T\sim T_{min}$) for all three cases, i.e., $\b=0,\pm1$.

\begin{table}[H]
    \begin{center}
   
\renewcommand{\arraystretch}{1.7}
\begin{tabular}{|c|c|c|c|c|}
 \hline
 $\beta$ &    $\s_{thermal}^{nM}/\s_{thermal}^{fM} $  & $\s_{U(1)}^{nM}/\s_{U(1)}^{fM}  $ &$ \sigma^{nM}_{U(1)}/\sigma_{thermal}^{nM}$ & $\sigma_{U(1)}^{fM}/\sigma_{thermal}^{fM}$
 \\ 
 \hline
 $1$ &$\frac{9nE}{\pi^2\th^2T_{min}^2}>>1$
   & $ \frac{9\sqrt{3} En}{\pi ^2\theta^2 T_{min }^2}>>1$&$\frac{9\sqrt{3}J_t}{4\pi^2T_{min}^2 }>>1$ &$\frac{9J_t}{4\pi^2T_{min}^2 }>>1$\\
  \hline
 $0$ & $ \frac{9 E n }{2 \pi ^2 (\pi-\th) ^2T_{min}^2}>>1 $&  $  \frac{9\sqrt{3} E n}{2 \pi ^2 (\pi-\theta) ^2 T_{\min }^2}>>1$ &$\frac{9\sqrt{3}J_t}{4\pi^2T_{min}^2 }>>1$ &$\frac{9J_t}{4\pi^2T_{min}^2 }>>1$\\
 \hline
  $-1$ &$\frac{9nE}{\pi^2(\pi-\th)^2T_{min}^2}>>1$ &  $\frac{9 \sqrt{3} E n}{\pi ^2(\pi-\theta) ^2 T^2{}_{\min }}>>1$&$\frac{9\sqrt{3}J_t}{4\pi^2T_{min}^2 }>>1$ &$\frac{9J_t}{4\pi^2T_{min}^2 }>>1$\\ \hline
\end{tabular}
 \caption{The ratio of conductivities for all three cases $\b=0,\pm1$ in the low temperature ($T\sim T_{min}$) regime upto leading order. Here, $nM$ and $fM$ denote points that are near and far from the Misner string, respectively.}
    \label{tableltn}
         
    \end{center}
\end{table}

\subsection{High temperature regime}
In this Section, we investigate $\s_{U(1)}$ (\ref{UTT}) and $\s_{thermal}$ (\ref{ETT}) in the high temperature regime for all three cases $\b=0,\pm1$. We define the high temperature domain as the regime where the temperature is significantly higher than the minimum temperature (\ref{miniT}), i.e., $T>>T_{min}$. 

To begin with, we first examine the case for \(\beta = 1\), where the Misner string is located at \(\theta = 0\). In the high temperature regime, $\s_{U(1)}$ (\ref{UTT}) and $\s_{thermal}$ (\ref{ETT}) can be expressed  both near and far from the Misner string as follows
\begin{align}
  &\sigma_{U(1)}\bigg|^{fM}_{T>>T_{min}}\approx\frac{9  J_t}{4 \pi (3+9n^2)}\left(\frac{T_{min}}{T}\right)^2,\label{highu11aa}\\&\sigma_{U(1)}\bigg|_{T>>T_{min}}^{nM}\approx\frac{9  J_t}{4 \pi (3+9n^2)}\left(\frac{T_{min}}{T}\right)^2+\frac{2187 E^2 n^2  J_t}{8 (3+9n^2)^3\theta ^4}\left(\frac{T_{min}}{T}\right)^6,\label{highu11} \\
&\sigma_{thermal}\Bigg|^{fM}_{T>>T_{min}}\approx1\hspace{1mm},\hspace{2mm}\sigma_{thermal}\Bigg|^{nM}_{T>>T_{min}}\approx1+\frac{81E^2 n^2 }{2(3 + 9 n^2)^2 \theta^4 }\left(\frac{T_{min}}{T}\right)^4.\label{highth1}
\end{align}
%\textcolor{red}{\begin{align}
 % &\sigma_{U(1)}\bigg|^{fM}_{T>>T_{min}}\approx\frac{9  J_t}{16 \pi ^2T^2},\sigma_{U(1)}\bigg|_{T>>T_{min}}^{nM}\approx\frac{9 J_t}{16 \pi ^2 T^2}+\frac{2187 E^2 n^2  J_t}{8192 \pi ^6 T^6}\csc ^4\left(\frac{\theta }{2}\right)\label{highu11} \\
   %&\sigma_{thermal}\Bigg|^{fM}_{T>>T_{min}}\approx1\hspace{1mm},\hspace{2mm}\sigma_{thermal}\Bigg|^{nM}_{T>>T_{min}}\approx1+\frac{81 E^2 n^2 }{512 \pi ^4  T^4}\csc ^4\left(\frac{\theta }{2}\right).\label{highth1}
%\end{align}}
The ratio of conductivities for the near Misner ($nM$) string and far away ($fM$) from it can be expressed as
\begin{align}
    \frac{\sigma_{U(1)}\Bigg|_{T >> T_{min}}^{nM} }{\sigma_{U(1)}\Bigg|_{T >> T_{min}}^{fM} }=1+O\left[\left(\frac{T_{min}}{T}\right)^4\right]\hspace{1mm},\hspace{2mm}\frac{\sigma_{thermal}\Bigg|_{T >> T_{min}}^{nM} }{\sigma_{thermal}\Bigg|_{T >> T_{min}}^{fM} }=1+O\left[\left(\frac{T_{min}}{T}\right)^4\right].\label{highratiob1}
\end{align}
%{\textcolor{red}{\begin{align}
    %\frac{\sigma_{U(1)}\Bigg|_{T >> T_{min}}^{nM} }{\sigma_{U(1)}\Bigg|_{T >> T_{min}}^{fM} }=1+O(1/T^4)\hspace{1mm},\hspace{2mm}\frac{\sigma_{thermal}\Bigg|_{T >> T_{min}}^{nM} }{\sigma_{thermal}\Bigg|_{T >> T_{min}}^{fM} }=1+O(1/T^4).\label{highratiob1}
%\end{align}}}

\begin{figure}[htp]
     \centering
     \begin{subfigure}[b]{0.495\textwidth}
         \centering
         \includegraphics[width=\textwidth]{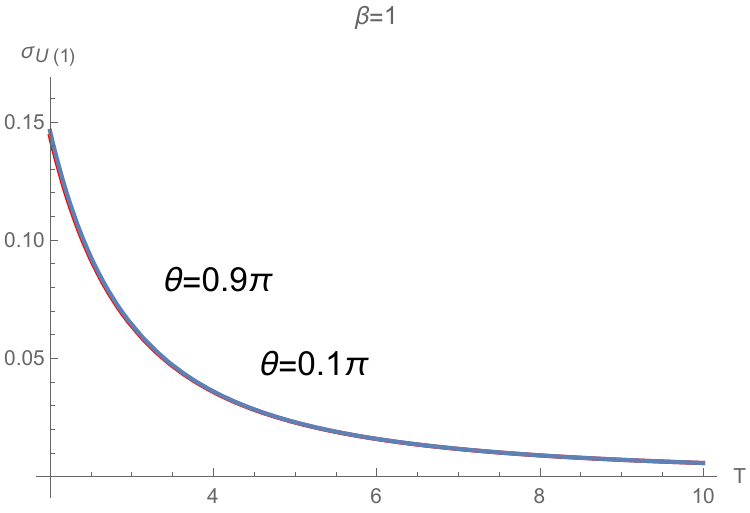}
         \caption{$\s_{U(1)}$ vs temperature}\label{FIG4A}
              \end{subfigure}
     \hfill
     \begin{subfigure}[b]{0.495\textwidth}
         \centering
         \includegraphics[width=\textwidth]{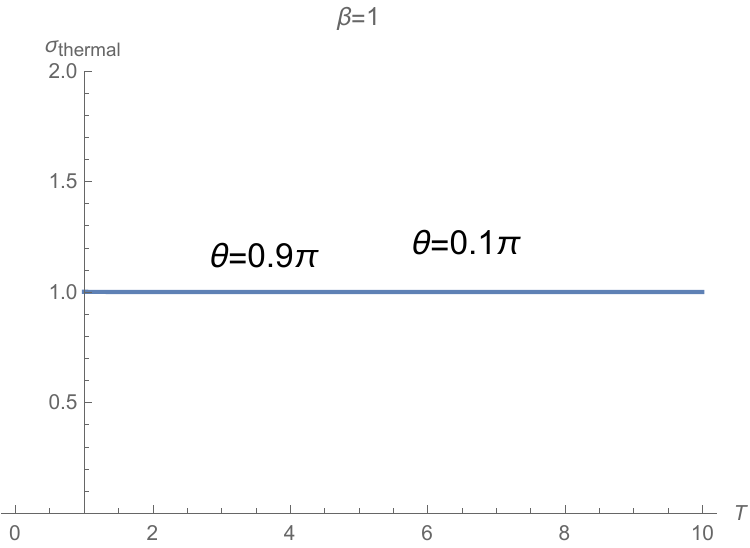}
         \caption{$\s_{thermal}$ vs temperature}\label{FIG4B}
             \end{subfigure}
             \caption{Conductivity vs temperature plot for $\b=1$. For points near the Misner string ($nM$), we set $\theta=0.1\pi$, and for points away from the Misner string ($fM$), we set $\theta=0.9\pi$. Here, we set $n=0.2$, $E=0.1$ and $J_t=10$.}
        \label{figure4}
\end{figure}

The above expressions (\ref{highu11aa})-(\ref{highth1}) show that the U(1) conductivity ($\s_{U(1)}$) falls with increase in the temperature \cite{Kim:2011zd}-\cite{Lee:2010ii}, while $\s_{thermal}$ remains nearly constant, regardless of the position of the Misner string as illustrated in Figure \ref{figure4}. Furthermore, equation (\ref{highratiob1}) indicates that at leading order in the expansion, the conductivity due to the U(1) charge carriers near the Misner string ($nM$) is almost the same as that of points located farther away from the Misner string ($fM$). This observation also applies to the conductivity of the thermally produced charge pairs. Consequently, it implies that the effects of frame dragging are suppressed at high temperatures\footnote{Notice that the angular velocity (\ref{av1})-(\ref{angb2}) in the high temperature ($T>>T_{min}$) limit goes as $|\tilde{\Omega}|\approx\frac{9 n|\cos\theta+\beta|}{8 \left(3 n^2+1\right)^2}\left(\frac{T_{min}}{T}\right)^4$. As a result, the frame dragging effects completely disappear in high temperature regimes.} for $\b=1$ (\ref{av1}).
  
\vspace{2mm}
$\bullet$ \textbf{For cases $\b=0$ and $\b=-1$}
\vspace{1mm}

Finally, one can conduct a similar analysis for the other two cases, namely, $\b=0$ and $\b=-1$.  For $\b=0$, the conductivity resulting from U(1) charge carriers (\ref{UTT}) and the thermally generated charge carriers (\ref{ETT}) near and far from the Misner string can be expressed as follows
\begin{align}
&\sigma_{U(1)}\bigg|^{fM}_{T>>T_{min}}\approx\frac{9  J_t}{4 \pi (3+9n^2)}\left(\frac{T_{min}}{T}\right)^2\hspace{1mm},\label{he0aa}\\&\sigma_{U(1)}\bigg|^{nM}_{T>>T_{min}}\approx\frac{9  J_t}{4 \pi (3+9n^2)}\left(\frac{T_{min}}{T}\right)^2+\frac{64J_t E^2 n^2  }{ (3 + 9 n^2)^3(\pi-\theta)^4}\left(\frac{T_{min}}{T}\right)^6,\label{he0}\\
   & \sigma_{thermal}\bigg|^{fM}_{T>>T_{min}}\approx1\hspace{1mm},\hspace{2mm}\sigma_{thermal}\bigg|^{nM}_{T>>T_{min}}\approx1+\frac{81E^2 n^2 }{8(3 + 9 n^2)^2 (\pi-\theta)^4 }\left(\frac{T_{min}}{T}\right)^4.\label{hth0} 
\end{align}
%{\textcolor{red}{\begin{align}
%&\sigma_{U(1)}\bigg|^{fM}_{T>>T_{min}}=\frac{9  J_t}{16 \pi ^2T^2},\sigma_{U(1)}\bigg|^{nM}_{T>>T_{min}}=\frac{9 J_t}{16 \pi ^2 T^2}+\frac{2187 e^2 n^2 \cot ^2(\theta ) \csc ^2(\theta ) J_t}{2048 \pi ^6 T^6}\label{he0}\\
 %  & \sigma_{thermal}\bigg|^{fM}_{T>>T_{min}}=1,\hspace{18mm}\sigma_{thermal}\bigg|^{nM}_{T>>T_{min}}=1+\frac{81 E^2 n^2 \cot ^2\theta  \csc ^2\theta }{128 \pi ^4 T^4}\label{hth0} 
%\end{align}}}

%Below, we plot $\s_{U(1)}$ (\ref{UTT}) and $\s_{thermal}$ (\ref{ETT}) against temperature for both cases, $\b=0$ and $\b=-1$, as shown in Figures \ref{figure5} and \ref{figure6}, respectively. For the case of \(\beta=0\), we set \(\theta=0.9\pi\) near the Misner string, while for points further away, we set \(\theta=0.4\pi\). On the other hand, for \(\beta=-1\), we set \(\theta=0.9\pi\) close to the Misner string and \(\theta=0.1\pi\) for points that are far away from it. In both cases, we set \(n=0.2\), $E=0.1$ and $J_t=10$.

Additionally, the ratio of conductivities (\ref{he0aa})-(\ref{hth0}) close to and away from the Misner string can be expressed as follows %\footnote{\textcolor{red}{The extra $\frac{1}{\epsilon^2}$ factor for $\beta=0$ case is not associated with frame dragging and nut parameter, it arises due to the global factor $\frac{1}{\sin^2\theta}$ in the conductivity e.i $\frac{\sin\th|_{\th \to \frac{\pi}{2}}}{\sin\th |_{\th \to \epsilon,\pi-\epsilon}}=\frac{1}{\epsilon^2}$}}
\begin{align}
   \frac{\sigma_{U(1)}\Bigg|_{T >> T_{min}}^{nM} }{\sigma_{U(1)}\Bigg|_{T >> T_{min}}^{fM} }=1+O\left[\left(\frac{T_{min}}{T}\right)^4\right]\hspace{1mm},\hspace{2mm}\frac{\sigma_{thermal}\Bigg|_{T >> T_{min}}^{nM} }{\sigma_{thermal}\Bigg|_{T >> T_{min}}^{fM} }=1+O\left[\left(\frac{T_{min}}{T}\right)^4\right].\label{highratiob0}  
\end{align}
%{\textcolor{red}{\begin{align}
 %   \frac{\sigma_{U(1)}\Bigg|_{T >> T_{min}}^{nM} }{\sigma_{U(1)}\Bigg|_{T >> T_{min}}^{fM} }=1+O(1/T^4)\hspace{1mm},\hspace{2mm}\frac{\sigma_{thermal}\Bigg|_{T >> T_{min}}^{nM} }{\sigma_{thermal}\Bigg|_{T >> T_{min}}^{fM} }=1+O(1/T^4).\label{highratiob1}
%\end{align}}}
%where $\epsilon<<1$ is a parameter (see footnote \ref{fn2}).

These expressions (\ref{he0aa})-(\ref{highratiob0}) are similar as found for $\b=1$ case. The only notable difference appears in the subleading terms (\ref{he0})-(\ref{hth0}), where $\th\sim\pi$ in the case of $nM$ string expansion.

%\textcolor{red}{rewrite...}It is important to note that when \(\beta = 0\), the Misner string is located at both \(\theta = 0\) and \(\theta = \pi\). If we choose to focus on one position of the Misner string (say, \(\theta = \pi\)), the point farthest from the Misner string is located at \(\theta = \frac{\pi}{2}\). When we consider the ratio of conductivities (\ref{highratiob0}), this results in an overall factor of \(1/\epsilon^2\). Consequently, \(\sigma_{U(1)}\) (or \(\sigma_{thermal}\)) is greater near the Misner string compared to the point farther away\footnote{The frame dragging effects are negligible in the high temperature regime (\ref{angb1}). }, as shown in Figure \ref{figure5}.
\begin{figure}[htp]
     \centering
     \begin{subfigure}[b]{0.495\textwidth}
         \centering
         \includegraphics[width=\textwidth]{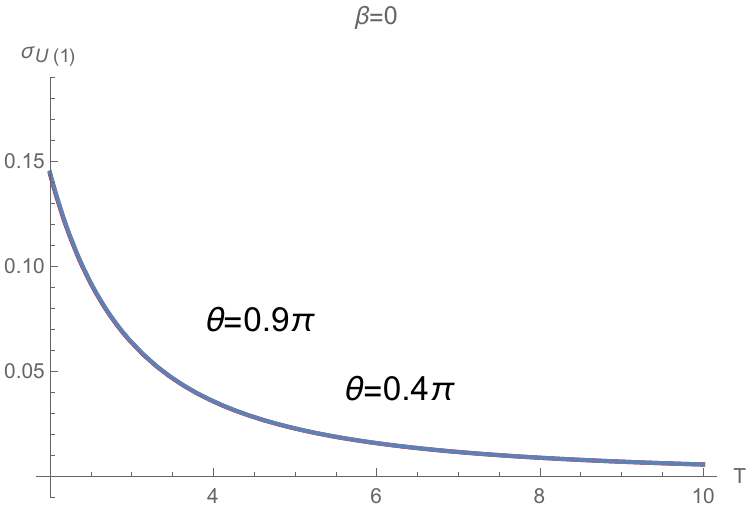}
         \caption{$\s_{U(1)}$ vs temperature}\label{FIG5A}
              \end{subfigure}
     \hfill
     \begin{subfigure}[b]{0.495\textwidth}
         \centering
         \includegraphics[width=\textwidth]{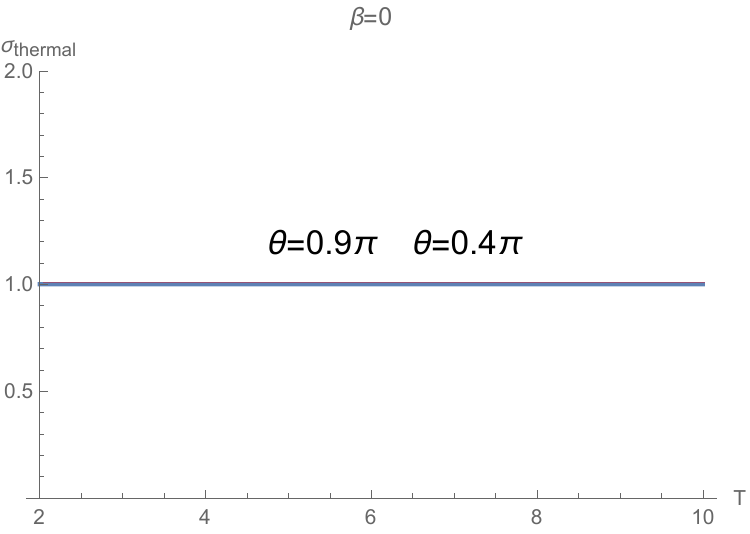}
         \caption{$\s_{thermal}$ vs temperature}\label{FIG5B}
             \end{subfigure}
             \caption{Conductivity vs temperature plot for $\b=0$. For points close to the Misner string, we set $\theta=0.9\pi$, and for points away from the Misner string, we set $\theta=0.4\pi$. Moreover, we set $n=0.2$, $E=0.1$ and $J_t=10$.}
        \label{figure5}
\end{figure}

%\textcolor{red}{do it in last at once...}Notably, the thermal conductivity (\ref{hth0}) dominates over the conductivity resulting from the U(1) charge carrier (\ref{he0}), as illustrated in Figure \ref{figure5}.  This phenomenon can be understood by the fact that, in the high temperature regime, the number of thermally generated neutral charge pairs exceeds the number of electric charge carriers.

Finally, we repeat the analysis for the case $\b=-1$. Both $\s_{U(1)}$ and $\s_{thermal}$, near and away from the Misner string, can be expressed as 
 \begin{align}
    &\sigma_{U(1)}\bigg|^{fM}_{T>>T_{min}}=\frac{9  J_t}{4 \pi (3+9n^2)}\left(\frac{T_{min}}{T}\right)^2,\label{hem1aa}\\&\sigma_{U(1)}\bigg|^{nM}_{T>>T_{min}}=\frac{9  J_t}{4 \pi (3+9n^2)}\left(\frac{T_{min}}{T}\right)^2+\frac{2187 E^2 n^2  J_t}{8 (3+9n^2)^3(\pi-\theta) ^4}\left(\frac{T_{min}}{T}\right)^6,\label{hem1}\\
   & \sigma_{thermal}\bigg|^{fM}_{T>>T_{min}}=1\hspace{1mm},\hspace{2mm}\sigma_{thermal}\bigg|^{nM}_{T>>T_{min}}=1+\frac{81E^2 n^2 }{2(3 + 9 n^2)^2 (\pi-\theta)^4 }\left(\frac{T_{min}}{T}\right)^4.\label{hthm2}
\end{align}
%{\textcolor{red}{\begin{align}
 %   &\sigma_{U(1)}\bigg|^{fM}_{T>>T_{min}}=\frac{9  J_t}{16 \pi ^2T^2}, \sigma_{U(1)}\bigg|^{nM}_{T>>T_{min}}=\frac{9 J_t}{16 \pi ^2 T^2}+\frac{2187 E^2 n^2  J_t}{8192 \pi ^6 T^6}\sec ^4\left(\frac{\theta }{2}\right),\label{hem1}\\
   %& \sigma_{thermal}\bigg|^{fM}_{T>>T_{min}}=1,\hspace{1mm}\sigma_{thermal}\bigg|^{nM}_{T>>T_{min}}=1+\frac{81 E^2 n^2 }{512 \pi ^4  T^4}\sec ^4\left(\frac{\theta }{2}\right)\label{hthm2}
%\end{align}}}
The ratio of conductivities reveals an identical expression as found before
\begin{align}
   \frac{\sigma_{U(1)}\Bigg|_{T >> T_{min}}^{nM} }{\sigma_{U(1)}\Bigg|_{T >> T_{min}}^{fM} }=1+O\left[\left(\frac{T_{min}}{T}\right)^4\right]\hspace{1mm},\hspace{2mm}\frac{\sigma_{thermal}\Bigg|_{T >> T_{min}}^{nM} }{\sigma_{thermal}\Bigg|_{T >> T_{min}}^{fM} }=1+O\left[\left(\frac{T_{min}}{T}\right)^4\right].\label{highratiobm1}
\end{align}
%{\textcolor{red}{\begin{align}
 %   \frac{\sigma_{U(1)}\Bigg|_{T >> T_{min}}^{nM} }{\sigma_{U(1)}\Bigg|_{T >> T_{min}}^{fM} }=1+O(1/T^4)\hspace{1mm},\hspace{2mm}\frac{\sigma_{thermal}\Bigg|_{T >> T_{min}}^{nM} }{\sigma_{thermal}\Bigg|_{T >> T_{min}}^{fM} }=1+O(1/T^4).\label{highratiob1}
%\end{align}}}
%where $\epsilon<<1$ is a parameter (see footnote \ref{fn2}).
%\newpage
\begin{figure}[htp]
     \centering
     \begin{subfigure}[b]{0.495\textwidth}
         \centering
         \includegraphics[width=\textwidth]{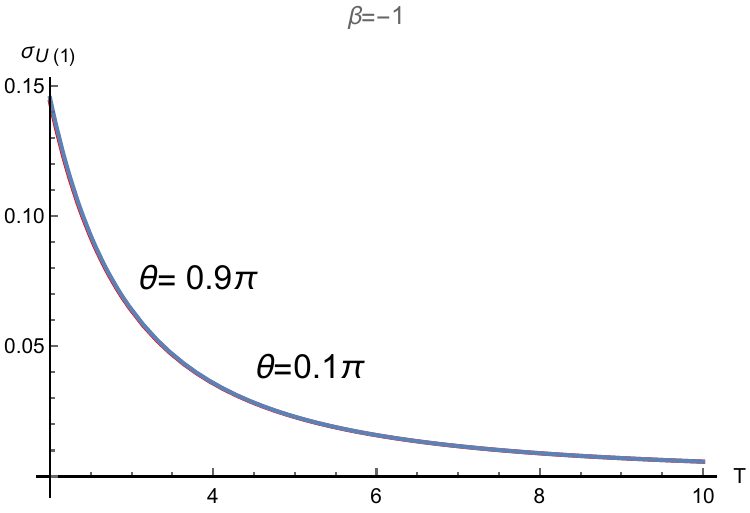}
         \caption{$\s_{U(1)}$ vs temperature}\label{FIG6A}
              \end{subfigure}
     \hfill
     \begin{subfigure}[b]{0.495\textwidth}
         \centering
         \includegraphics[width=\textwidth]{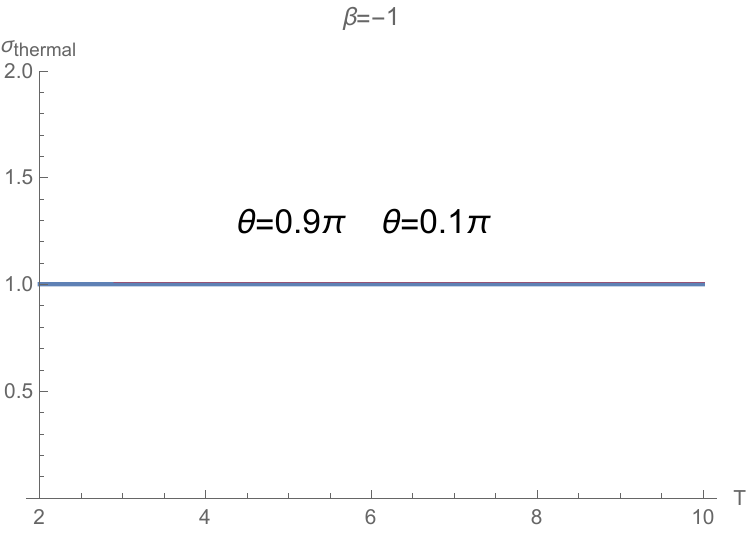}
         \caption{$\s_{thermal}$ vs temperature}\label{FIG6B}
             \end{subfigure}
             \caption{Conductivity vs temperature plot for $\b=-1$. For points near the Misner string, we set $\theta=0.9\pi$, and for farther points, we set $\theta=0.1\pi$. Further, we set $n=0.2$, $E=0.1$, and $J_t=10$.}
        \label{figure6}
\end{figure}

%Notice that the behaviors of $\s_{U(1)}$ (\ref{hem1aa})-(\ref{hem1}) and $\s_{thermal}$ (\ref{hthm2}) in the high temperature regime for $\b=-1$ are similar to those observed for $\b=0$ and $\b=1$. In particular, $\s_{U(1)}$ falls with rise in temperature, while  $\s_{thermal}$ remains almost constant,  both near and far from the Misner string, as illustrated in Figure \ref{figure6}. Notably, as mentioned before, in high-temperature regions, the effects of frame dragging are diminished (\ref{angb2}), resulting in the nearly same value of $\s_{U(1)}$ both near and far from the Misner string (\ref{highratiobm1}). A similar observation also applies to $\s_{thermal}$.

It is interesting to notice that in the large $T(>>T_{min})$ limit, $\s_{thermal}$ (for all three cases $\b=0,\pm1$), becomes independent of both the NUT parameter ($n$) and the temperature, regardless of the location of the Misner string. This observation is consistent with the fact that the thermal conductivity behaves as \(\s_{thermal} \sim T^{|p-2|/z}\) in the high temperature regime, where (\(p+2\)) represents the total number of space-time dimensions\footnote{For our case,  $p=2$.} \cite{Lee:2010uy}.

Furthermore, the thermal conductivity for all three cases $\b=0,\pm1$ dominates over the U(1) charge carriers (upto leading order) in the high temperature region (\ref{highu11aa})-(\ref{hthm2}), regardless of the position of the Misner string. This can be seen by taking ratios
\begin{align}
    \frac{\s_{U(1)}^{nM}}{\s_{thermal}^{nM}}\Bigg|_{\b=0,\pm1} =\frac{9  J_t}{4 \pi (3+9n^2)}\left(\frac{T_{min}}{T}\right)^2<<1,\\
    \frac{\s_{U(1)}^{fM}}{\s_{thermal}^{fM}}\Bigg|_{\b=0,\pm1}=\frac{9  J_t}{4 \pi (3+9n^2)}\left(\frac{T_{min}}{T}\right)^2<<1.
\end{align}
The above ratio clearly shows that the number of thermally produced charge pairs exceeds the U(1) charge carriers in the high temperature ($T>>T_{min}$) regime, leading to a larger thermal conductivity. This behavior is in contrast with that observed at low temperatures ($T\sim T_{min}$), where thermal conductivity is suppressed as compared to the U(1) conductivity (\ref{dcratio}).

%\textcolor{red}{do it in last at once...}Finally, the thermally produced charge carriers exceed the U(1) charge carriers in the high temperature regime, resulting in a greater thermal conductivity (\ref{hthm2}) both near and far from the Misner string as shown in Figure \ref{figure6}.

Below, in Table \ref{tableltnn}, we summarize the key findings for all three cases, namely $\b=0,\pm1$ in the high temperature ($T>> T_{min}$) regime.

%As a result, the entire contribution to resistivity (\ref{resist1},) originates from the thermally generated neutral charge carriers, which yields
%\begin{align}
 %   \rho\sim\rho_{thermal}\sim1,\label{rhigh}
%\end{align}
%where $\rho_{thermal}$ represents the thermal resistivity. 

%In the limit $T\rightarrow\infty$, $\s_{thermal}$ ($\sim1/\sin^2{\th}$) (\ref{highth1}) becomes independent of both the NUT charge and the temperature, independent of the location of the Misner string. This observation is consistent with the fact that the conductivity behaves as \(\s \sim T^{|p-2|/z}\) in the high temperature regime, where (\(p+2\)) represents the total number of space-time dimensions\footnote{For our case,  $p=2$.} \cite{Lee:2010uy}.

\begin{table}[H]
    \centering
\renewcommand{\arraystretch}{1.7}
\begin{tabular}{|c|c|c|c|c|}
 \hline
 $\beta$ &    $\s_{thermal}^{nM}/\s_{thermal}^{fM}  $  & $\s_{U(1)}^{nM}/\s_{U(1)}^{fM}  $ & $\sigma^{nM}_{U(1)}/\sigma_{thermal}^{nM}$&$\sigma^{fM}_{U(1)}/\sigma_{thermal}^{fM}$
 \\ 
 \hline
 $1$ &$ \approx1$
   & $\approx1$&$\frac{9  J_t}{4 \pi (3+9n^2)}\left(\frac{T_{min}}{T}\right)^2<<1$&$\frac{9  J_t}{4 \pi (3+9n^2)}\left(\frac{T_{min}}{T}\right)^2<<1$\\
  \hline
 $0$ & $\approx 1$&  $\approx1$&$\frac{9  J_t}{4 \pi (3+9n^2)}\left(\frac{T_{min}}{T}\right)^2<<1$&$\frac{9  J_t}{4 \pi (3+9n^2)}\left(\frac{T_{min}}{T}\right)^2<<1$\\
 \hline
  $-1$ &$\approx1$ &  $\approx1$&$\frac{9  J_t}{4 \pi (3+9n^2)}\left(\frac{T_{min}}{T}\right)^2<<1$&$\frac{9  J_t}{4 \pi (3+9n^2)}\left(\frac{T_{min}}{T}\right)^2<<1$\\ \hline
\end{tabular}
 \caption{The ratio of conductivities for all three cases $\b=0,\pm1$ in the high temperature regime.}
    \label{tableltnn}
\end{table}
%\begin{table}[H]
 %   \centering
%\renewcommand{\arraystretch}{1.7}
%\begin{tabular}{|c|c|c|c|c|}
% \hline
% $\beta$ &    $\sigma_{U(1)}/\sigma_{thermal}\Big|_{\text{nM}}^{\text{HT}}$  & $\sigma_{U(1)}/\sigma_{thermal}\Big|_{\text{fM}}^{\text{HT}}$ & $\rho\bigg|^{HT}_{nm} \approx\rho_{thermal}\Big|_{\text{nM}}^{\text{HT}}$ & $\rho\bigg|^{HT}_{nm} \approx\rho_{thermal}\Big|_{\text{fM}}^{\text{HT}}$
 %\\ 
 %\hline
 %$1$ &$\frac{9 \sqrt{n EJ_t}}{16\sqrt{2}\pi^2 T^2}<<1$
 %  & $\frac{9J_t}{4\pi^2 L^4T^2}<<1$& $1$ & $1$\\
  %\hline
 %$0$ &$\frac{9 \sqrt{n EJ_t}}{16\sqrt{2}\pi^2 T^2}<<1$&  $ \frac{9J_t}{4\pi^2 L^4T^2}<<1$&$1$& $1$ \\
 %\hline
  %$-1$ &$\frac{9 \sqrt{n EJ_t}}{16\sqrt{2}\pi^2 T^2}<<1$ &  $\frac{9J_t}{4\pi^2L^4 T^2}<<1$& $1$&$1$\\ \hline
%\end{tabular}
 %\caption{Conductivity and resistivity for various values for $\beta$ in the high temperature (HT) regime. Here, nM refers to near the Misner string, while fM refers to far from the Misner string.}
  %  \label{tablelt HT}
%\end{table}

%\newpage
\section{Conclusion and future direction}\label{cncbody}

To conclude, we explore the metallic holography in the context of the Taub-NUT AdS black hole in four space-time dimensions. In particular, we compute the associated holographic DC conductivity following the probe D-barne approach and study the effects of the NUT parameter ($n$) and the associated frame dragging on it. We treat the externally applied electric field ($E$) as a perturbation ($E<<1$). 

Our analysis reveals that the holographic DC conductivity consists of two parts. The first part ($\s_{U(1)}$) is due to the added charge carriers by the D-brane, while the other contribution ($\s_{thermal}$) arises due to the thermally produced charge pairs. We find that both types of conductivities are enhanced due to the presence of the NUT parameter ($n$), which in turn causes a frame dragging effect near the Misner string at low temperatures ($T\sim T_{min}$).

At low temperatures ($T\sim T_{min}$), we observe that the number of $U(1)$ charge carriers exceeds thermally produced charge pairs, leading to higher $U(1)$ conductivity. In contrast, far from the Misner string, only the $U(1)$ conductivity rises, while $\sigma_{thermal}$ remains constant. To summarise, at low temperatures, the increase in conductivity is sharper for points closer to the Misner string as compared to those farther away from it. We identify this as an artifact of larger drift of charge carriers near the Misner string.

It is interesting to notice that the U(1) conductivity at low temperatures ($T\sim T_{min}$) grows as $\s_{U(1)}\sim T^{-2}$ when we are away from the Misner string. Notably, the authors in \cite{Ge:2016lyn} reported a similar result, where they compute the conductivity associated with a black hole in the asymptotic Lifshitz space-time. These observations suggest that in the low temperature regime, the conductivity behaves like that of a Fermi liquid when we are sufficiently distant from the Misner string \cite{Ge:2015fmu}, \cite{Lee:2010uy}. Conversely, as we approach closer to the Misner string, frame dragging begins to dominate, leading to a conductivity behavior, \(\sigma_{U(1)} \sim T^{-4}\). This can be interpreted as a new phase of a quantum liquid \cite{Ge:2016lyn}.

Finally, in the high temperature regime ($T>>T_{min}$), we find that the number of thermally produced pairs is significantly higher than the added $U(1)$ charge carriers, resulting in a larger thermal conductivity. Interestingly enough, the frame dragging effects are quite suppressed in the high temperature regime. As a result, both the conductivities $\s_{U(1)}$ and $\s_{thermal}$ are least affected due to the presence of the Misner string. Furthermore, the thermal conductivity in the high temperature regime becomes independent of both the NUT parameter $(n)$ and temperature $(T)$, i.e. $\sigma_{thermal} \sim 1$. This observation is consistent with the result of \cite{Lee:2010uy} for a four-dimensional space-time.

Below, we outline some interesting future projects that are worth exploring.

$\bullet$ The authors in \cite{OBannon:2007cex} further generalized the approach of computing the holographic DC conductivity \cite{Karch:2007pd} and obtained the holographic Hall conductivity associated with the $AdS_5$ black hole. Therefore, it would be an interesting project to explore the holographic Hall conductivity in the context of the Taub-NUT AdS black hole and study the effects of the NUT parameter ($n$) and the associated frame dragging.

$\bullet$ Finally, it would be interesting to compute the holographic entanglement entropy within the framework of the Taub-NUT AdS$_4$ black holes using the Ryu-Takayanagi formula\footnote{The authors in \cite{Tavakoli:2018xnh} have conducted a similar study in the context of charged accelerating AdS black holes.} \cite{Ryu:2006ef} and explore the effects of NUT parameter $(n)$ on it.

We hope to address some of these issues in our future publications.

\section*{Acknowledgments}
AK and DR are indebted to the authorities of Indian Institute of Technology, Roorkee for their unconditional support towards researches in basic sciences. HR would like to thank the authorities of Saha Institute of Nuclear Physics, Kolkata, for their support. The authors would like to thank Aditya Mehra for useful discussions. DR acknowledges the Mathematical Research Impact Centric Support (MATRICS) grant (MTR/2023/000005) received from ANRF, India.

%\appendix
%\section{Current density and charge density}\label{Ab}
%\section{Exact expressions of $\s_{U(1)}$ and $\s_{thermal}$}\label{EXPP}
%In this appendix, we provide the complete expressions of $\sigma_{U(1)}$ and $\sigma_{thermal}$ as functions of temperature $T$. 

%Using (\ref{sig1})-(\ref{sig2}) and (\ref{miniT}), one finds
%\begin{align}
%&\s_{thermal}= \hspace{1mm}\frac{1}{\sin^2\theta}\sqrt{1+\frac{324 E^2 n^2 \csc ^4\theta  (\beta +\cos \theta )^2}{\left(4 \pi ^2  \left(2 T \left(\tilde{T}+T\right)-T_{\min }^2\right)+9 n^2\right)^2}}\\
%&\s_{U(1)}= \hspace{1mm}\frac{9 \csc ^2\theta  J_t}{\left(4 \pi ^2  \left(2 T \tilde{T}-T_{\min }^2+2 T^2\right)+9 n^2\right)}\Bigg[\frac{972 E^2 n^2 \csc ^4\theta  (\beta +\cos \theta )^2}{\left(4 \pi ^2  \left(2 T \left(\tilde{T}+T\right)-T_{\min }^2\right)+9 n^2\right)^2}
%\nonumber\\
%&-\frac{432 \pi ^2 e^2 \csc ^2\theta \left(\tilde{T}+T\right){}^2}{\left(8 \pi ^2 T \left(\tilde{T}+T\right)-4 \pi ^2 T_{\min }^2+9 n^2\right)^2 \left(8 \pi ^2 T \left(\tilde{T}+T\right)-4 \pi ^2 T_{\min }^2+9 n^2+3\right)}\nonumber\\
%& +\frac{4 E n \csc ^2\theta  (\beta +\cos \theta ) \left(81 J_t^2+\left(4 \pi ^2  \left(\tilde{T}+T\right)^2+9 n^2\right)^2\right)^{1/2}}{J_t \left(8 \pi ^2  T \left(\tilde{T}+T\right)-4 \pi ^2  T_{\min }^2+9 n^2\right)}\hspace{2mm}+1\Bigg]^{1/2},
%\end{align}
%where we define $\tilde{T}=\sqrt{T^2-T_{\min }^2}$.

\end{document}